\documentclass[preprint2]{aastex}

\shorttitle{Andromeda~XXI}
\shortauthors{Cusano et al.}

\begin{document}

\title{VARIABLE STARS AND STELLAR POPULATIONS IN ANDROMEDA~XXI: II. ANOTHER MERGED GALAXY SATELLITE OF M31?\altaffilmark{*}}

\author{FELICE CUSANO\altaffilmark{1}, ALESSIA GAROFALO\altaffilmark{1,2}, GISELLA CLEMENTINI\altaffilmark{1},
MICHELE CIGNONI\altaffilmark{3}, LUCIANA FEDERICI\altaffilmark{1}, 
MARCELLA MARCONI\altaffilmark{4}, ILARIA MUSELLA\altaffilmark{4}, VINCENZO RIPEPI\altaffilmark{4},
ROBERTO SPEZIALI\altaffilmark{5}, ELEONORA SANI\altaffilmark{6,7},
ROBERTO MERIGHI\altaffilmark{1}}

\affil{$^1$INAF- Osservatorio Astronomico di Bologna, Via Ranzani 1, I - 40127 Bologna, Italy}
\email{felice.cusano@oabo.inaf.it}
\affil{$^2$ The Observatories of the Carnegie Institution of Washington, 813 Santa Barbara Street, Pasadena, CA 91101, USA}
\affil{$^3$ Space Telescope Science Institute, 3700 San Martin Drive, Baltimore, MD 21218, USA}
\affil{$^4$INAF- Osservatorio Astronomico di Capodimonte, Salita Moiariello 16, 
I - 80131 Napoli, Italy}
\affil{$^5$INAF- Osservatorio Astronomico di Roma, Via di Frascati 33
00040 Monte Porzio Catone, Italy}
\affil{$^6$INAF - Osservatorio Astronomico di Arcetri, Largo Enrico Fermi 5, I - 50125 Firenze }
\affil{$^7$ESO, Alonso de Cordova 3107, Casilla 19001, Vitacura, Santiago 19, Chile}

\altaffiltext{*}{Based on data collected  with the Large Binocular Cameras at the Large Binocular Telescope}

\begin{abstract}
\noindent $B$ and $V$ time-series photometry of the M31 dwarf spheroidal satellite  Andromeda~XXI (And~XXI) was obtained 
with the Large Binocular Cameras at the Large Binocular Telescope. We have identified 50 variables in And~XXI, 
of which  41 are RR Lyrae stars (37 fundamental-mode $-$ RRab, and 4 first-overtone $-$RRc, pulsators) and 9 are Anomalous Cepheids (ACs).
The average period of the RRab stars ($\langle Pab \rangle$ = 0.64 days) and the period-amplitude diagram  place  And~XXI in the 
class of Oosterhoff  II - Oosterhoff-Intermediate objects. 
From the average luminosity of the RR Lyrae stars we derived the galaxy distance modulus of  (m-M)$_0$=$24.40\pm0.17$ mag,
which is smaller than previous literature estimates, although still consistent with them within 1 $\sigma$.
The galaxy color-magnitude diagram shows evidence for the presence of three different stellar generations in And~XXI: 
1) an old ($\sim$ 12 Gyr) and metal poor ([Fe/H]=$-$1.7 dex) component traced 
by the  RR Lyrae stars; 2) a slightly younger (10-6 Gyr)  and more metal rich ([Fe/H]=$-$1.5 dex) component populating the red horizontal branch, 
and 3) a young age ($\sim$ 1 Gyr) component with same metallicity, that produced the ACs. Finally,  
we provide hints that And~XXI could be the result of a minor merging event between two dwarf galaxies.
\end{abstract}


\section{INTRODUCTION}\label{sec:intro}
This is the second in our series of papers devoted to the study of  Andromeda's  (M31) satellites based on 
time-series photometry obtained with the Large Binocular Cameras (LBC) of  the Large Binocular Telescope (LBT). 
Our aim is to  characterize 
the resolved stellar populations of the M31 companions using the color-magnitude diagram (CMD) and the properties   
of  variable stars \citep[see][for a general description of the project]{cle11}. The final  goal is  to derive hints on 
the formation history of  Andromeda's  satellites and relate them to
the global context of merging and accretion episodes occurring in  M31.
Details of the survey and results from the study of  the M31 satellite Andromeda~XIX (And~XIX) were presented in \citet[][Paper~I]{cus2013}. In this paper we  
report  on the study of another very extended M31  dwarf spheroidal (dSph) companion:  Andromeda~XXI (And~XXI).
The satellite galaxies of M31  are indeed an important test-bed for  both
the $\Lambda$Cold Dark Matter ($\Lambda$CDM) and the Modified Newtonian Dynamics \citep[MOND,][]{mil83} theories.
In the  $\Lambda$CDM scenario the majority of the satellites are believed to be primordial dwarf
galaxies residing in dark matter halos, which merge and are accreted 
to form larger galaxies \citep{zol2009}. The 
satellites that we see today around M31 could thus be the residual building blocks of the M31 assembling process. 
However, some issues challenge the $\Lambda$CDM scenario, among them is
the phase-space distribution of the satellites around M31. \citet{iba2013}
discovered that 15 of  the more than 30 satellites of Andromeda lie in a thin plane 
12 kpc thick and over 200 kpc in size, which 
rotates around M31.  
Accretion of small satellites through filaments has been
invoked to explain the disky distribution of these M31 companions. However, the tiny width of the disk and the large  number of satellites in it  are features very hard to
reproduce by  $\Lambda$CDM simulations \citep[see][ for a complete discussion]{paw2014}.
A different explanation for the origin of the vast thin plane of satellites orbiting M31 is
that they are tidal dwarf galaxies formed in the debris of a past major merger between
M31 and a massive galaxy \citep{ham2013}. The space distribution of
the satellites is well reproduced by numerical models of galaxy-interaction.
A further alternative to the merger model is  that the satellites formed during 
 a past fly-by of the Milky Way (MW) and M31 about 10 Gyr ago. Indeed, under the assumption of
Milgromiam dynamics\footnote{\citet{mil83}}, \citet{zhao2013} found that M31 and the MW had a close
fly-by encounter between 7 and 11 Gyr ago. 
Understanding the real nature/origin of the M31 satellites is thus crucial to address
also cosmological theories.

\begin{table*}
\begin{center}
\caption[]{Log of And~XXI observations}
\label{tab:obs}
\begin{tabular}{l c c c c }
\hline
\hline
\noalign{\smallskip}
   Dates                 & {\rm Filter}  & N   & Exposure time &  {\rm Seeing (FWHM)}    \\
 	                 &		 &     &	      (s)       &    {\rm (arcsec)}\\
\noalign{\smallskip}
\hline
\noalign{\smallskip}
  October   8-11, 2010     &   $B$      & 6  & 420    &  1.4\\
  December  1-3, 2010  &   $B$      & 40 & 420    & 0.8  \\ 
                         &             &   &        &        \\  
  October   8-11, 2010   &   $V$      &  7 &   420  &  1.4 \\
  December  1-3, 2010   &   $V$      & 39 &  420  &  0.8   \\
\hline
\end{tabular}
\end{center}
\normalsize
\end{table*}
And~XXI  
was discovered by \citet{mart09} in the context of the PAndAS survey \citep[][ and reference therein]{mart13}.
The discovery paper reports a distance modulus 
of (m-M)$_0$= $24.67\pm0.13$ mag from the luminosity of the red giant branch tip and a half-light radius (r$_h$) of 3.5$'$ (corresponding to $r_h = 875 \pm 127$ pc at the distance of And~XXI).
This makes And~XXI the fourth largest dSph in the Local Group (LG). 
The galaxy has a metallicity of [Fe/H]=$-1.8\pm0.4$ dex, as estimated by \citet{col13} from the calcium triplet  (CaII) of the galaxy red giants. The same authors
measured a velocity dispersion  of $\sigma=4.5^{+1.2}_{-1.0}$ kms$^{-1}$  from 32 spectroscopically confirmed members,  
which is rather low, when compared with the great extension of And~XXI. This was interpreted by \citet{col13}  as the result of 
a possible past tidal interaction with M31, however,  for \citet{mcgaugh2013} 
the low dispersion naturally arises in the MOND context. 
\citet{con12} adopting a bayesian approach to locate the galaxy  RGB tip  have revised 
And~XXI  distance modulus to (m-M)$_0$= $24.59^{+0.06}_{-0.07}$ mag which, although shorter, is still within 
1 $\sigma$  of  \citet{mart09}'s value. 

The paper is organized as follows. Observations, data reduction and calibration of And~XXI photometry are presented in Section 2. 
Results on the identification and characterization of the variable stars, 
the catalog of light curves, and the Oosterhoff classification of And~XXI are discussed in  Sections 3. 
The distance to And~XXI derived from the  RR Lyrae stars is presented in Section 4.
The galaxy CMD is discussed in Section 5 and the  projected distribution of And~XXI stellar components is Section 6.  
Section 7 gives some hints on the possibility of a past merging. Finally, a summary of the main results is presented in Section 8.

\section{OBSERVATIONS AND DATA REDUCTION}

Time series photometry in the  $B$ and $V$ bands of a 23$^{\prime} \times 23^{\prime}$ region centered on And~XXI  (R.A.$=23^{h}54^{m}47.7^{s}$,
Decl. $=+42^{\circ}28^{\shortmid}15^{\shortparallel}$, J2000.0; \citealt{mart09}) was obtained in October and December 2010 with the LBC
at the foci of the LBT. Sub-arcsec seeing conditions were achieved  for the December observations, while seeing was slightly worse during the October run.
Observations in the $B$ band  were obtained with the Blue camera of the  LBC, whereas the $V$ images were acquired with the Red camera.
A total of  46 $B$ and 46 $V$ images each of  420s exposure were obtained for a total exposure time of  19320s  in each band. 
Images in both bands were dithered 
by 30 arcsec in order to fully cover the inter-CCD gaps of the LBC mosaic.
Observations of And~XIX were obtained in the same nights by interchanging the two targets in order to evenly sample the light curves of 
variable stars possibly occurring in the two galaxies.
The log of the observations of And~XXI is provided in Table~\ref{tab:obs}.
Data reduction was performed in the same way as for And~XIX and is described in Paper~I,  to which the interested 
reader is referred for details. The  
PSF photometry was performed using  the \texttt{DAOPHOT-ALLSTAR-ALLFRAME} package  \citep{ste87,ste94}.
Since the observations of And~XXI were acquired within a few minutes from those of And~XIX,  for the absolute photometric calibration we 
used the calibrating equations derived in Paper~I, by properly accounting for differences in airmass between the two targets.

\section{VARIABLE STARS}
Identification of the variable stars was performed using the variability index computed in \texttt{DAOMASTER} \citep{ste94}, then 
the light curves of the candidate variables were analyzed
with the Graphical Analyzer of Time Series (GRATIS), custom software developed at the Bologna Observatory 
by P. Montegriffo \citep[see, e.g.,][]{clm00}. Further details on the search for variables and the analysis of the light curves
can be found in Paper~I.
A total of 50 variable stars were identified both in the $B$ and $V$ band datasets.
The  properties of the variable stars detected in And~XXI are summarized in Table ~\ref{t:1}. 
We  named the variables with an increasing number starting from the galaxy center, for which we adopted the coordinates by \citet{mart09}. 
Column 1 gives the star identifier, Columns 2 and 3 provide the right ascension and declination (J2000 epoch), respectively. 
These coordinates were obtained from our astrometrized catalogs. Column 4 gives the type of variability. 
Columns 5 and 6 list the period and the Heliocentric Julian Day of maximum light, respectively.  
Columns 7 and 8 give the intensity-weighted mean $B$ and $V$ magnitudes, while Columns 9 and 10 list the corresponding amplitudes of the light variation.
Light curves are presented in Figure~\ref{fig:lca}. 

\begin{table*}
\caption[]{Identification and properties of the variable stars detected in And~XXI}
\footnotesize
\label{t:1}
\begin{tabular}{l c c l l c c c c c }
\hline
\hline
\noalign{\smallskip}
 Name & $\alpha$            &$\delta$       & Type & ~~~P        & Epoch (max)& $\langle B \rangle$ & $\langle V \rangle$ & A$_{B}$ & A$_{V}$ \\ 
 	    &	(2000)   & (2000)   &	         &~(days)& JD ($-$2455000) & (mag)                       & (mag)         & (mag)     &  (mag)     \\
 	    \noalign{\smallskip}
	    \hline
	    \noalign{\smallskip}
	    
V1  &      23:54:47.066     & +42:28:19.42  & RRc  &    0.3950   & 533.684 &  25.87            &   25.38               &  0.66  & 0.51 \\   
V2  &      23:54:47.688     &   +42:27:16.38  & RRab &  0.6085  &  533.658 & 25.96             & 25.57               &  1.00 &   1.29\\
V3  &      23:54:49.205    & +42:29:15.29   & RRab &  0.5810     &  477.692 & 25.77            & 25.30             &  1.21   & 0.95  \\
V4 &     23:54:46.450      & +42:27:09.43   & AC   &     1.2353   &  532.625 &  24.22          &    23.63     &   1.33   & 1.03 \\
V5 &    23:54:46.788       &  +42:29:24.59  & RRab &  0.6708       & 533.608  & 25.74        & 25.22          &  0.99     & 0.77 \\
V6 &   23:54:51.778        & +42:27:37.58  & AC    &  1.067    &     476.959 &  24.63        &   24.07       &   1.06     &   0.53\\ 
V7 &  23:54:49.066        &  +42:29:24.25  & RRab &  0.6132      &     532.665  &  25.84      &  25.38         & 1.11    &  0.78  \\ 
V8 &      23:54:46.471    &  +42:29:37.67  &  RRab &  0.8342     &    531.621  & 25.55     &  24.93           & 0.83    &  0.8  \\
V9 & 23:54:42.230         &  +42:27:49.18 & RRc   &  0.3869      &    532.718 & 26.02     &  25.52          & 0.60    & 0.37    \\
V10 &   23:54:53.642       &   +42:27:45.65   & RRab &    0.6371    &   532.605 & 25.82      & 25.33         &  0.87 & 0.57  \\ 
V11 & 23:54:39.698         &   +42:29:12.73   & RRab &  0.6210    &     533.745 & 25.82      & 25.21& 0.96 & 0.81\\
V12 & 23:54:49.212        &   +42:30:26.64   & RRab  & 0.6662   &     533.727   & 25.85    & 25.31  & 1.25  &  1.06  \\
V13 &  23:54:40.649    &      +42:26:34.55&  AC  &    1.138    &   533.754    &   24.53  &  24.09 & 1.66  & 1.34 \\
V14 &23:54:38.774   & +42:29:26.20        & AC   & 1.072  &       533.590 & 24.96  &24.44  &  1.18 &  0.85  \\
V15 &23:54:52.061  & +42:30:34.56       &  AC  & 1.715  &    477.670   & 24.15 & 23.63  & 1.27 & 0.99 \\ 
V16 &   23:54:38.664    &+42:26:52.66    &   RRab & 0.643     &     533.510         & 25.78  & 25.31  & 1.30 & 1.15    \\ 
V17 &   23:54:56.498     &   +42:30:06.08 & RRab &  0.5768     &  532.663             & 25.78 & 25.26 &1.30 &  1.08 \\
V18 & 23:54:36.979      &  +42:29:18.24 & RRab  &  0.5785&  533.691  & 25.80 & 25.30 & 1.25 & 1.00 \\
V19 &  23:54:36.562      &    +42:27:21.82 &  AC   & 1.390 & 531.670  & 24.27 & 23.80 & 1.32 &  1.00   \\
V20 & 23:54:52.507       &  +42:30:56.59 & RRab   &  0.6193 & 532.710 & 25.99 & 25.41 & 0.83 & 0.64  \\
V21 &  23:54:37.603      &   +42:30:16.38 & RRab &0.5971 & 533.717 & 25.84  & 25.32  &  1.19 & 0.92  \\
V22 & 23:54:54.931      & +42:31:04.19  & RRab    & 0.5911 & 532.740&  25.87 & 25.36 & 0.78  & 0.63   \\
V23 &  23:54:35.311 &     +42:29:34.87    & RRab & 0.6661 &  531.686 &  25.74 & 25.21 & 1.24 & 1.05 \\
V24 & 23:54:59.700  &  +42:26:26.59       & RRab & 0.6019 & 531.690   &  25.87 &  25.37 & 1.08 & 0.79 \\
V25& 23:54:33.816 & +42:27:37.80          & RRab & 0.6368 &  531.750  &   25.93 &  25.42 & 0.86 & 0.76  \\
V26 & 23:55:01.913 & +42:28:29.10       & AC & 1.1500 &   480.518 & 24.05  &  23.58   & 1.19  & 0.94 \\
V27 & 23:55:01.992 & +42:28:35.47      &  RRab &  0.7052 & 480.660 & 25.60 & 25.04   & 0.59 & 0.46   \\
V28 &  23:54:59.134 & +42:26:03.77     & RRc  &  0.3058  &   533.711  & 25.68 & 25.34 & 0.76 & 0.56  \\
V29 & 23:55:01.531 & +42:27:10.69      & RRab &  0.6414  &   532.605  & 25.78 & 25.28 & 0.78 & 0.60   \\
V30 & 23:54:45.857 & +42:24:38.95     & RRab &  0.5898   &    531.768  & 25.74 &  25.28& 1.33 & 1.12 \\
V31 & 23:55:02.167 & +42:27:34.31     & RRab  &  0.5880   &  533.640      & 25.74 & 25.31 &  0.85 & 0.80  \\
V32 &  23:55:01.817 & +42:29:23.64    & RRab & 0.6513   &   533.530     &  25.79  & 25.38 & 1.12 & 0.86  \\
V33 & 23:54:52.793  & +42:24:41.00   & RRab &  0.6182  &  533.750         & 25.95 &25.41 &1.20 & 0.89 \\
V34 &   23:54:39.821 & +42:31:39.79   & AC   &  1.2460   &   532.702       & 24.45 & 24.01 & 1.30 &0.98 \\
V35 & 23:54:44.129  & +42:32:06.32  & RRab   & 0.6061  &  532.770        &  25.92 & 25.31 & 1.13 & 0.88  \\
V36 & 23:54:47.352  & +42:32:16.44  & RRab &  0.6167   &  480.650        &  25.88 & 25.37 & 0.94 & 0.77  \\
V37 & 23:54:34.003  &   +42:30:25.91 &   RRab & 0.7074  & 532.768       & 25.77 & 25.26 &  0.84 &0.79   \\
V38 &   23:54:57.665 & +42:31:48.01 & RRc   & 0.2851 &   532.700        & 25.89 & 25.45  & 0.63  &0.39   \\
V39 &  23:54:35.669  & +42:31:30.14 & RRab   & 0.5707  &    477.705   &  25.87 & 25.38  & 1.02  & 0.91  \\
V40 &  23:54:47.770  & +42:23:10.75 & RRab    & 0.7178 &    532.760  & 25.77 & 25.28 &  0.82   & 0.64  \\
V41 &   23:54:45.545 & +42:33:18.90   &  RRab    & 0.5860 &   531.715   & 25.89 & 25.45 &  1.24   & 0.87   \\
V42 &    23:54:39.228  &  +42:23:22.13   & RRab    &  0.6824 &  531.748  &  25.70 & 25.21  & 1.17  &   0.85   \\
V43 &     23:54:36.629 &   +42:33:24.98  & RRab    & 0.6078  &  531.747 &  25.88 & 25.35 & 1.16  & 0.91  \\
V44 & 23:55:04.075    &  +42:32:50.46   &AC    &   1.1655   &  477.405  &  24.21  & 23.82 & 1.12   &  0.87 \\
V45 &   23:54:43.080  &   +42:21:30.78   & RRab  &  0.5670   &  533.650     &  25.80   & 25.34 & 1.17 &  1.03 \\
V46 &   23:55:14.446  &   +42:31:01.42   & RRab   & 0.71:    & 533.726      & 25.89    & 25.30  & 0.81 &  0.63 \\
V47 &    23:55:16.555 & +42:28:51.39     &RRab   &  0.5646  &  532.655     &25.84  &  25.34    & 1.51  & 1.17   \\
V48 & 23:55:21.782  &   +42:29:46.14  &  RRab & 0.7560   &  532.795 &  25.83 & 25.36    & 0.68 & 0.61  \\
V49 &     23:53:54.898 &   +42:25:56.04 & RRab   & 0.6766   &    532.652      &  25.79 & 25.29 &   0.94  & 0.73  \\
V50 &  23:55:46.274    &   +42:26:55.86    & RRab    & 0.6920  &    531.762       &  25.84 &  25.39 &  0.80 &  0.57  \\
\hline 
\end{tabular}
\normalsize
\end{table*}

\begin{figure*}
\centering
\includegraphics[width=14cm,height=16cm]{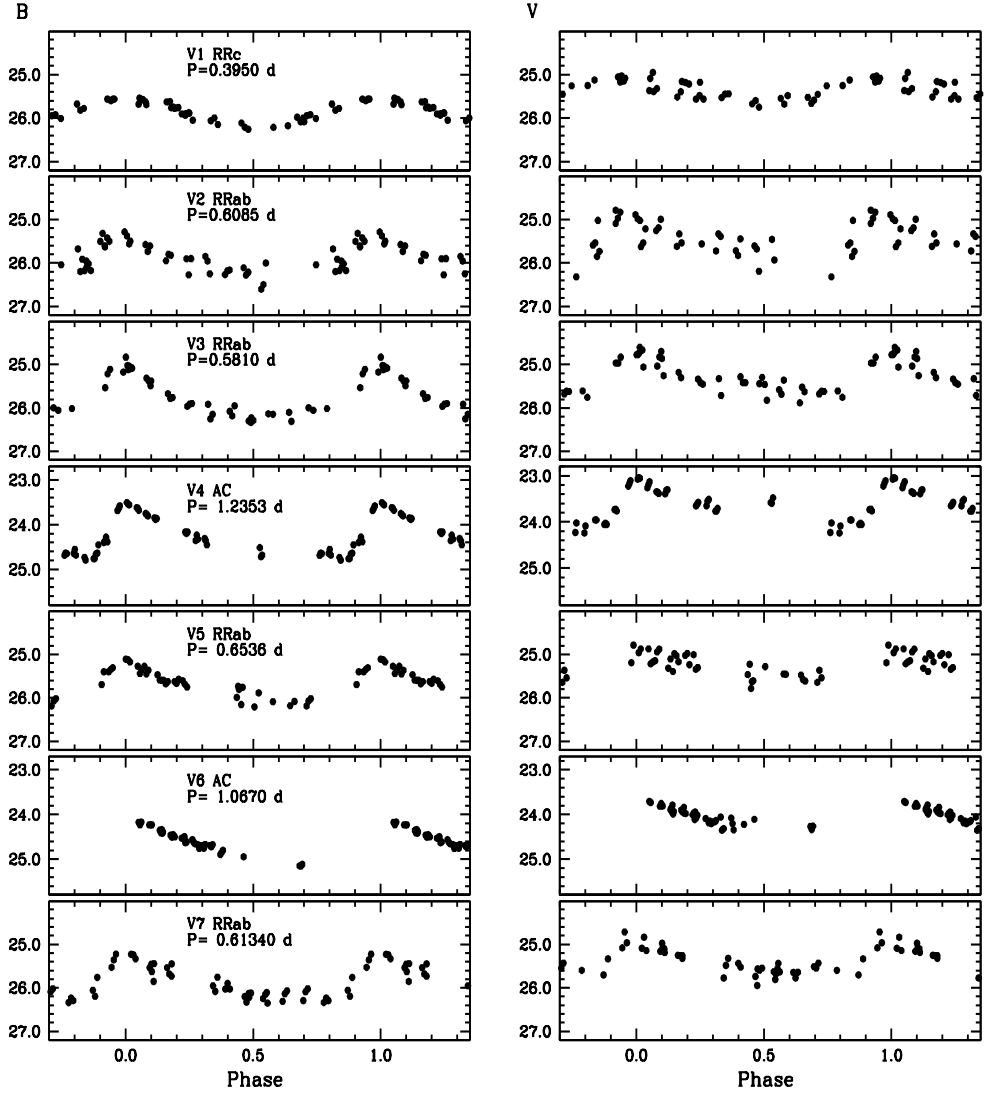}
\caption{$B$ (\textit{left panels}) and $V$ (\textit{right panels}) light curves of the variable stars identified  in And~XXI. 
Stars are ordered with increasing the distance from the galaxy  center, for which we adopted \cite{mart09} coordinates.
Typical internal errors for the single-epoch data are in the range of 0.02 to  0.18 mag in $B$, and of  0.03 to  0.25 mag in $V$.}
\label{fig:lca}
\end{figure*}
\begin{figure*}
\centering
\figurenum{1}
\includegraphics[width=14cm,height=16cm]{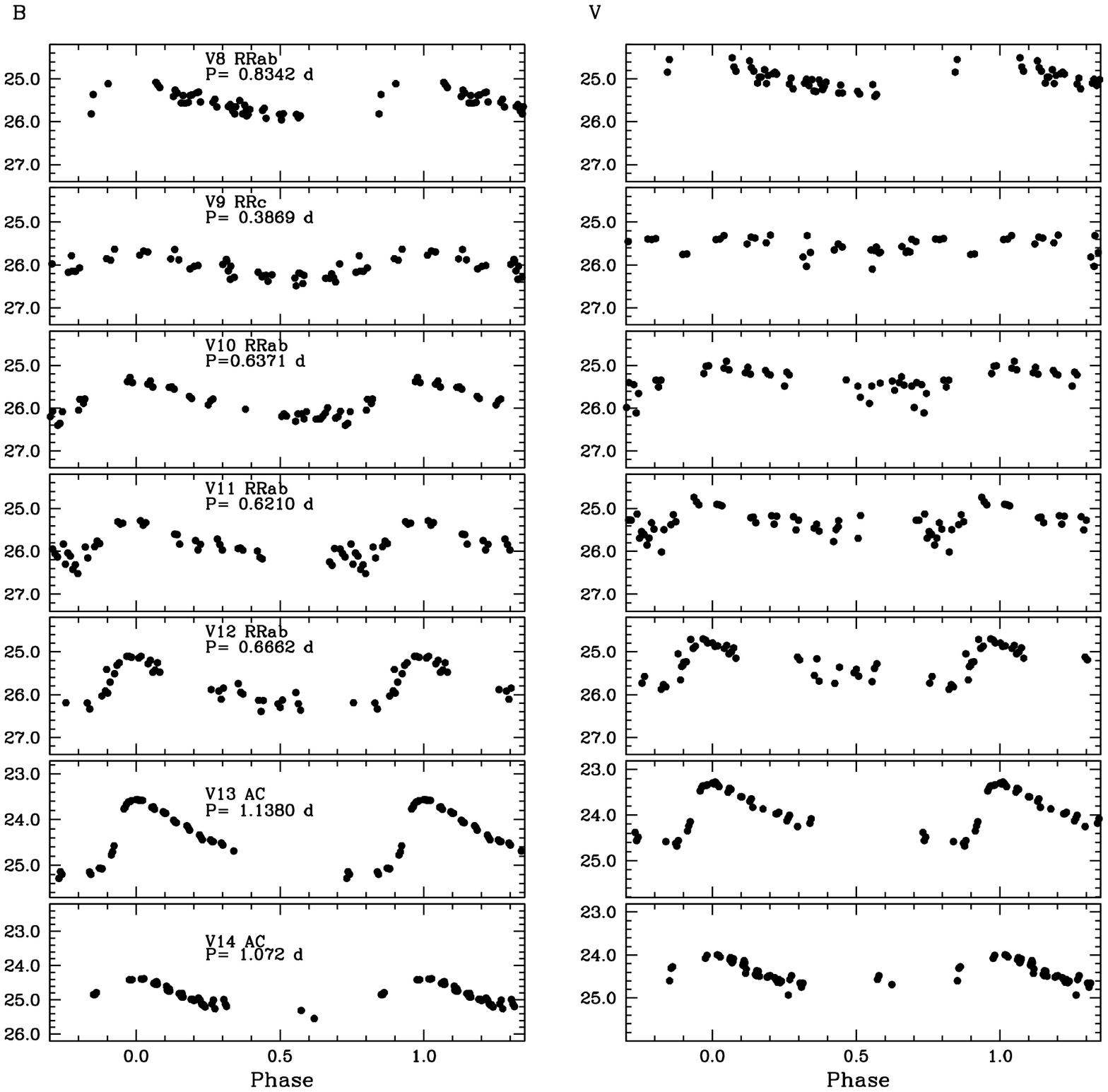}
\caption{ continued --}
\end{figure*}
\begin{figure*}
\centering
\figurenum{1}
\includegraphics[width=14cm,height=16cm]{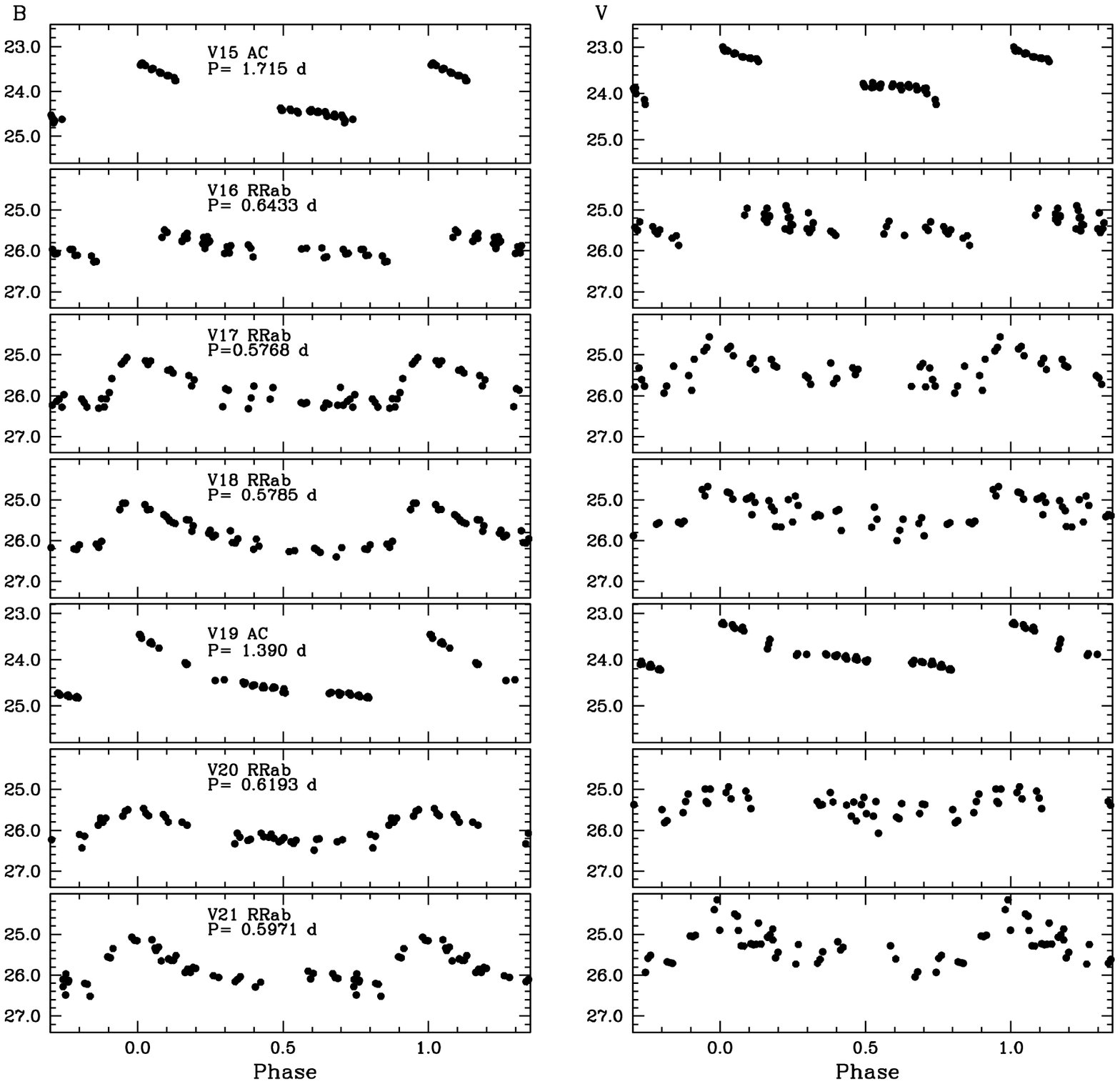}
\caption{ continued --}
\end{figure*}
\begin{figure*}
\centering
\figurenum{1}
\includegraphics[width=14cm,height=16cm]{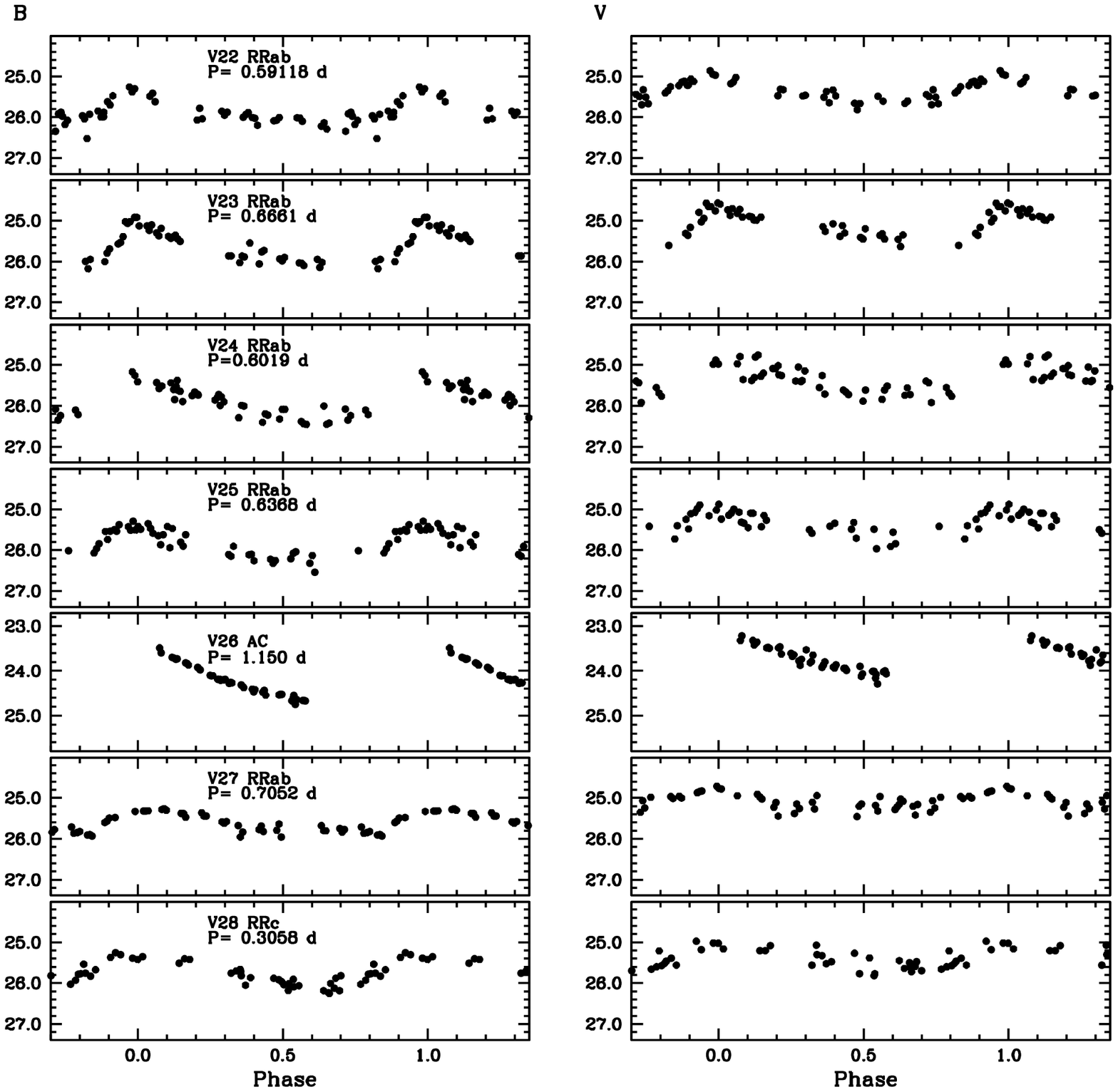}
\caption{ continued --}
\end{figure*}
\begin{figure*}
\centering
\figurenum{1}
\includegraphics[width=14cm,height=16cm]{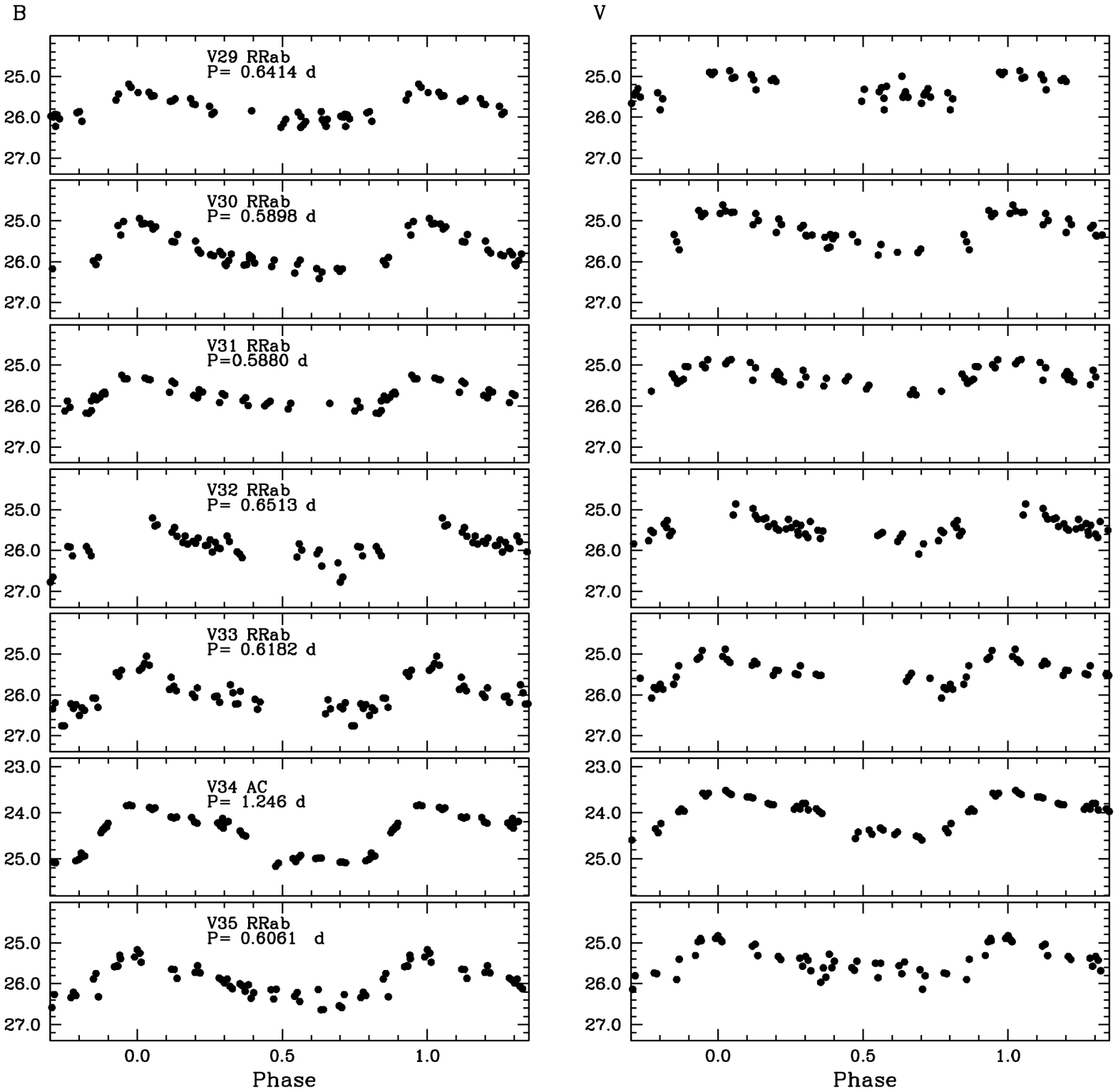}
\caption{ continued --}
\end{figure*}
\begin{figure*}
\centering
\figurenum{1}
\includegraphics[width=14cm,height=16cm]{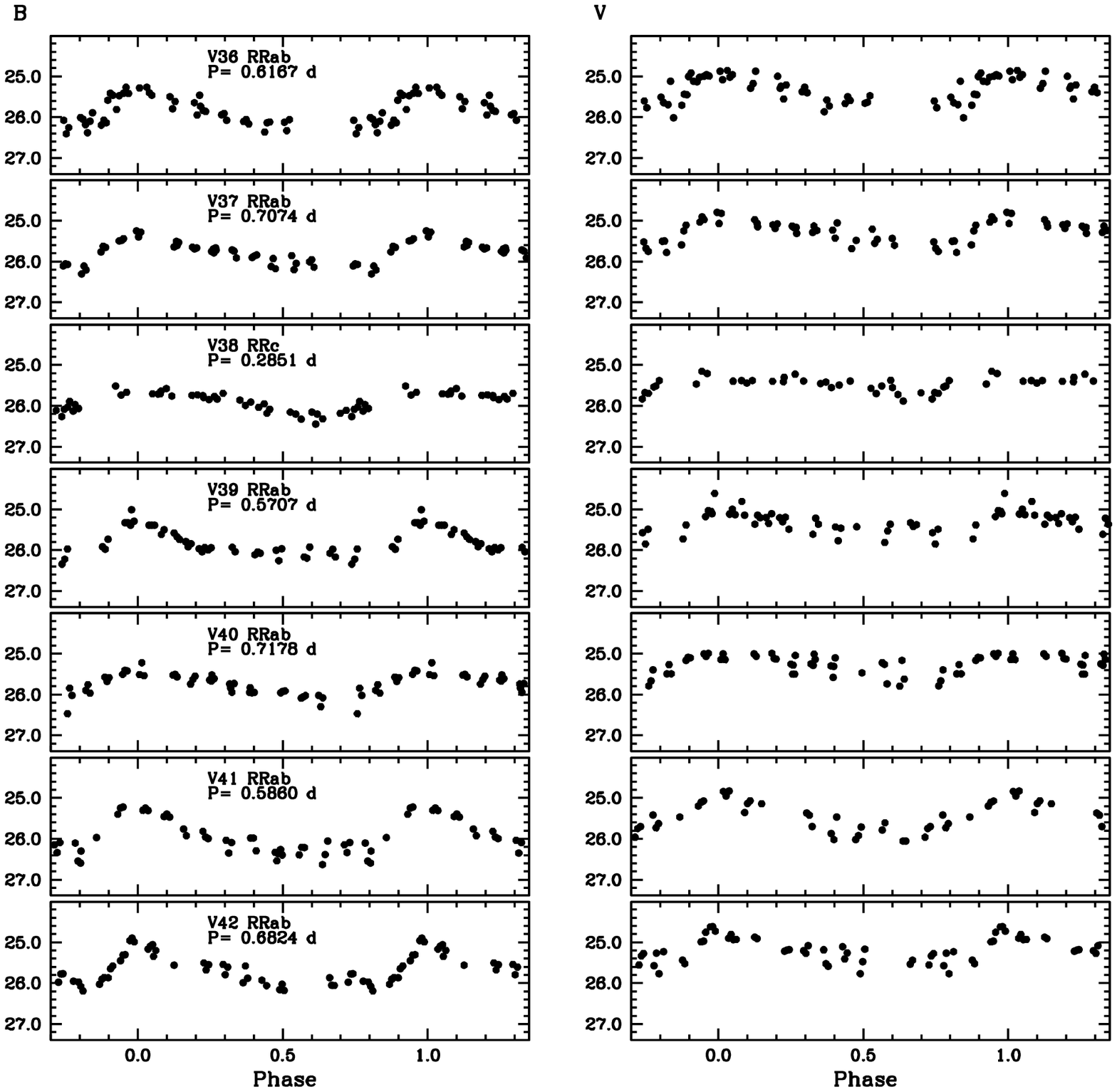}
\caption{ continued --}
\end{figure*}
\begin{figure*}
\centering
\figurenum{1}
\includegraphics[width=14cm,height=16cm]{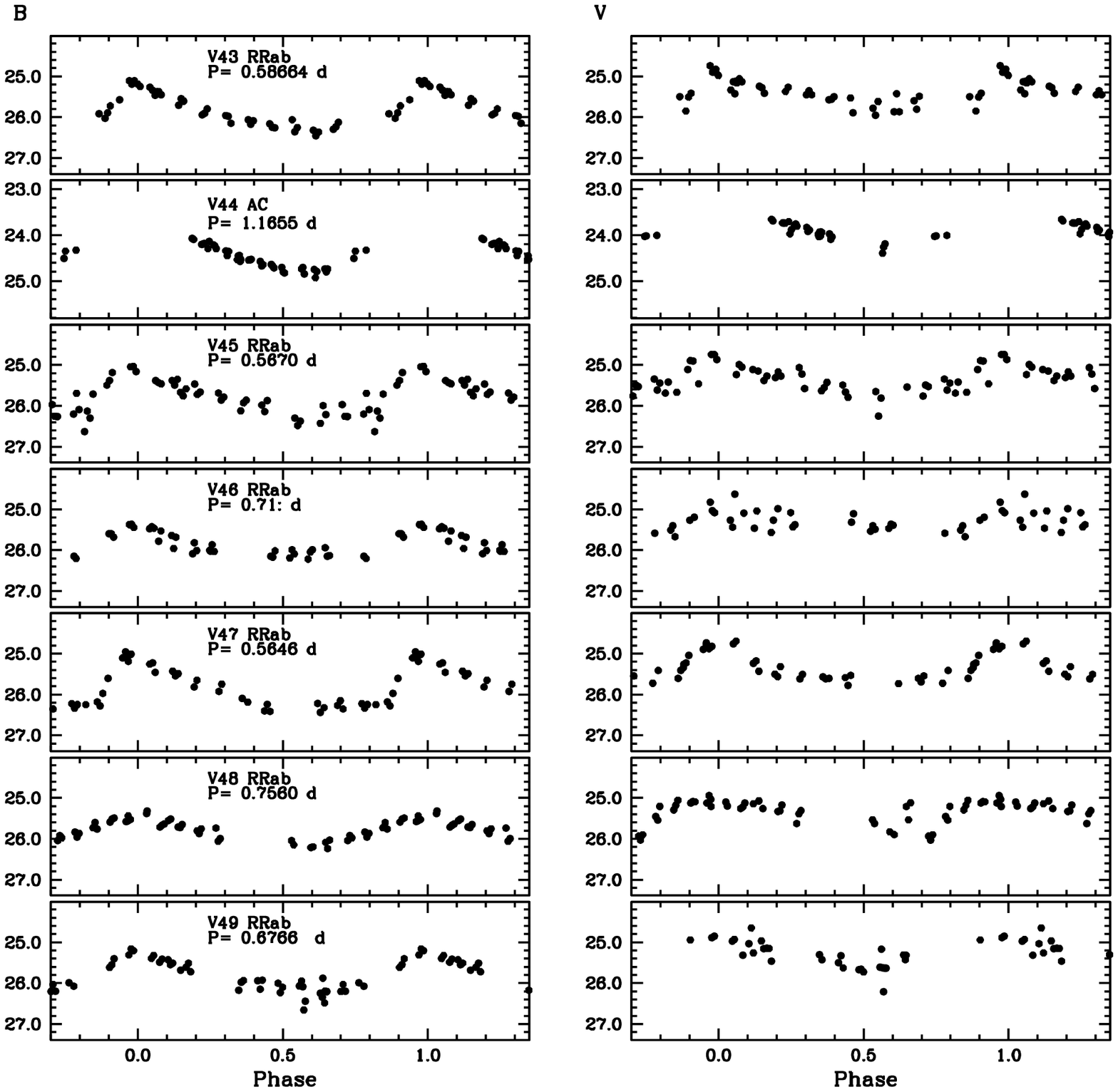}
\caption{ continued --}
\end{figure*}
\begin{figure*}
\centering
\figurenum{1}
\includegraphics[trim=0.001mm 11cm 0.001mm 0.001cm, keepaspectratio=true, width=14cm,height=16cm]{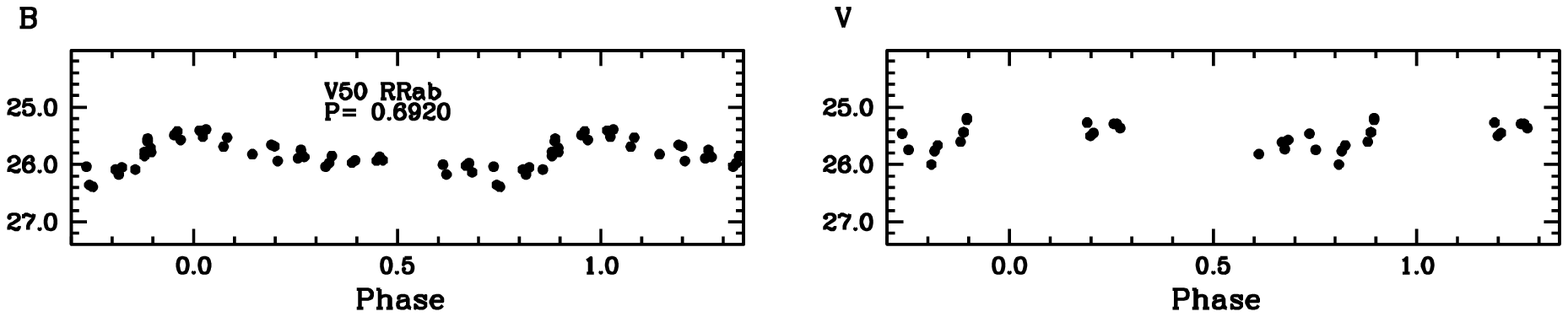}
\caption{ continued --}
\end{figure*}

\subsection{RR Lyrae stars}\label{sec:rrli}

\begin{figure}[b!]
\includegraphics[width=8.0cm,clip]{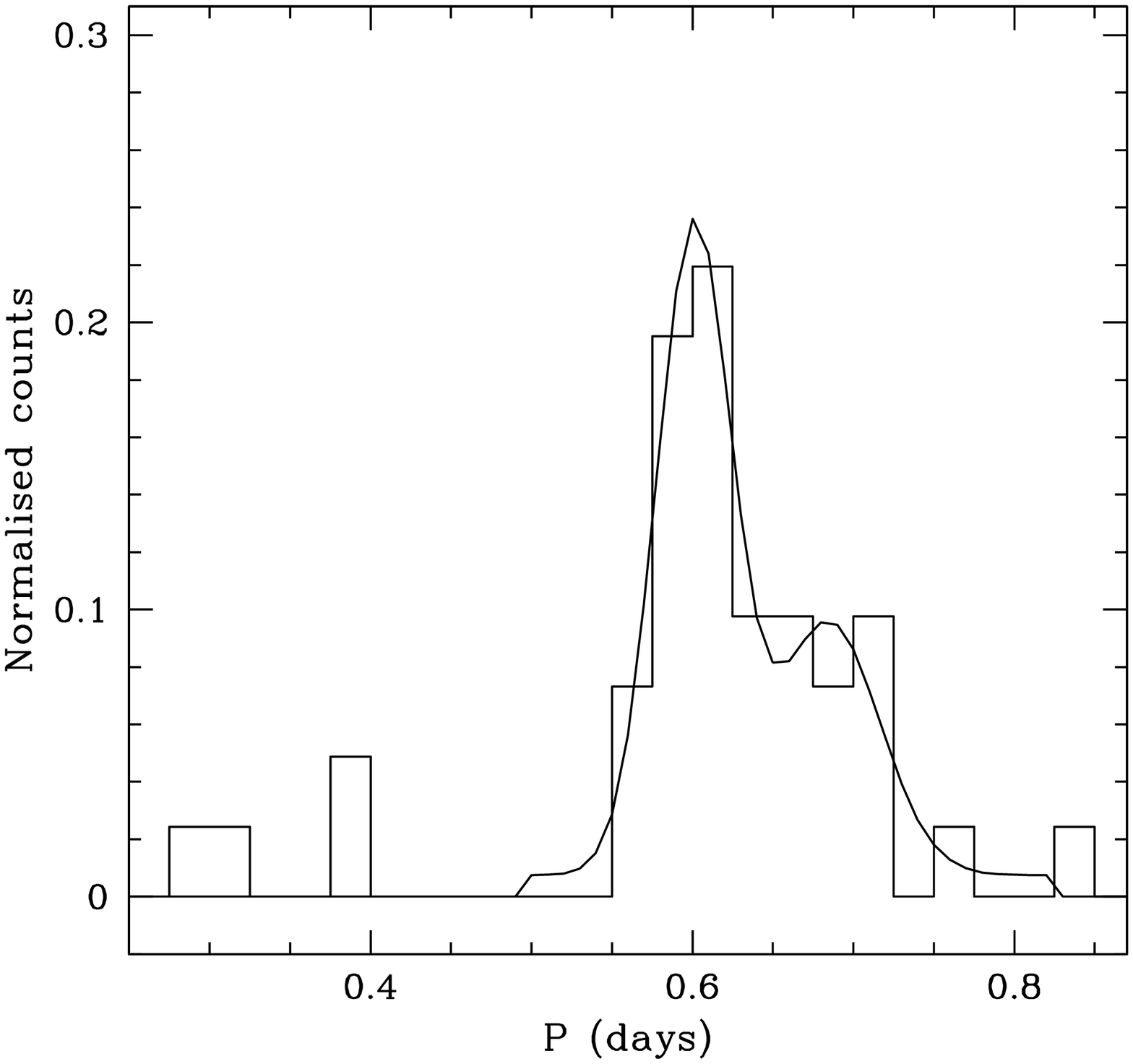}
\caption[]{Histogram of the periods of  the RR Lyrae stars identified in And~XXI. 
The bin size is 0.025 days. The gaussian fit to the period histogram is also shown.
The two stars at P=0.8342 d and P=0.756 d are V8 and V48, respectively.}
\label{fig:hist}
\end{figure}

We discovered a total of 41 RR Lyrae stars in And~XXI, of which 37 are RRab and 4 RRc pulsators. 
The average period of the 37 bona fide RRab stars is $\langle$P$_{\rm ab}\rangle$=0.64 d ($\sigma$=0.06 d). Considering 
only RRab  stars inside the galaxy half-light radius the average period becomes $\langle$P$_{\rm ab}\rangle$=0.63 days ($\sigma$=0.05 days, average on 22 stars).
From the  average period of the RRab stars, And~XXI can be classified as an Oosterhoff  II (Oo~II)/ Osterhoff Intermediate (Oo~Int) object (\citealt{oos39}).
However, since the r.m.s of the average period is rather large, in order to better investigate the Oosterhoff nature of And~XXI 
we produced  the histogram of the RR Lyrae periods. This is shown in 
Figure~\ref{fig:hist}. A multi-Gaussian  fit was performed to 
 find
the peak value of this distribution.  
We found two separate maxima, a first higher peak is at P$_1$=0.60 d with $\sigma=0.02$ d
and contains about 36 \%  of the RR Lyrae stars, and a second one 
at P$_2$=0.68 d with $\sigma=0.03$ d and about  24 \%  of the variables. The fit is shown in Figure~\ref{fig:hist}.
 The presence and relative strength of these two peaks reinforces for And~XXI a classification as Oo~Int/Oo~II  object.
In the histogram two RR Lyrae stars, V8 and V48, show  longer periods when compared to the average distribution. 
We comment on these stars further in Section~\ref{sec:comments}.

\begin{figure}[t!]
\includegraphics[width=7.7cm,clip]{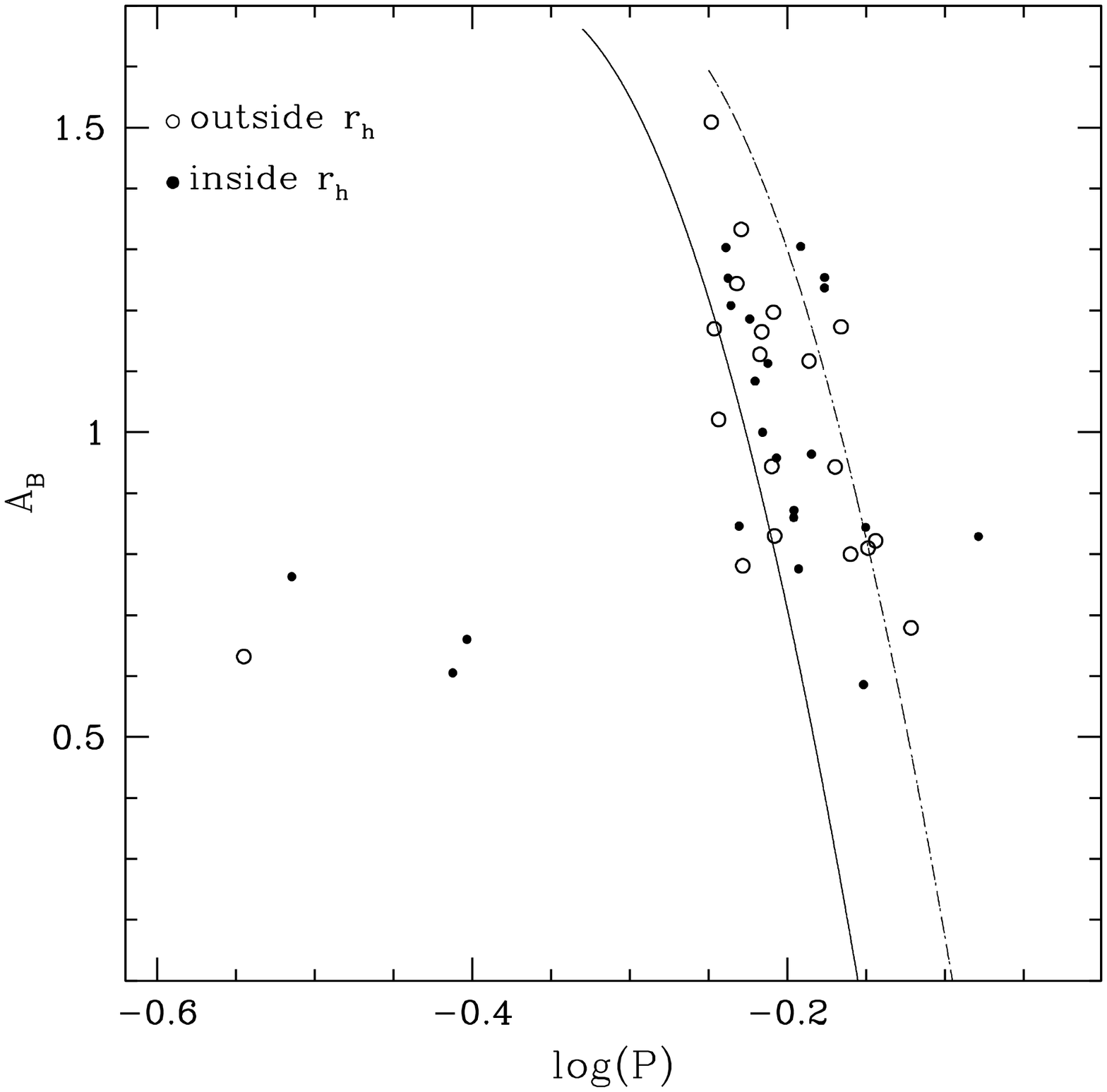}
\caption[]{ Period-amplitude diagram for the RR Lyrae stars in And~XXI. Solid and dashed lines represent the loci defined by  
bona-fide regular and well-evolved  RR Lyrae stars in M3 (Cacciari et al. 2005), respectively.  Filled  and open symbols mark variables within and outside the galaxy half-light radius, respectively.
}
\label{fig:bayl}
\end{figure}

The period-amplitude diagram (also known as Bailey diagram, \citealt{Bai1902}) of the And~XXI RR Lyrae stars is shown in Figure~\ref{fig:bayl} together
with the loci defined by  
bona-fide regular (solid line) and the well-evolved  (dashed line) RR Lyrae stars in the Galactic globular cluster M3,  
according to \citet{cac05}. M3 regular RR Lyrae stars have Oosterhoff I (Oo~I) properties, while the M3 evolved variables mimic the location of the Oo~II variables. 
The majority of the And~XXI RR Lyrae stars fall between the two loci and near the position of the Oo~II locus, 
thus confirming the galaxy classification as an Oo~Int/Oo~II object.
The fraction of RRc stars over the total number of RR Lyrae stars is f$_c$=N$_c$/N$_{\rm ab+c}$=0.10, 
that is small both for an Oo~II  (f$_c\sim0.44$) 
 and an Oo~I (f$_c\sim0.17$) system. This fraction is also small compared to other dwarf galaxies 
of similar metallicity \citep[Fe/H=-1.8 dex, ][]{col13} like for example And~XIX (f$_c$=0.26, Paper~I).
RRc stars  have smaller amplitudes, hence they are more difficult to identify than RRab stars. Furthermore, they have bluer colors.
Hence, as it will be discussed in Section~5, they occur in the region of the 
CMD which is heavily contaminated by background galaxies. This could affect our capability of detecting RRc stars. However, 
we individually checked for variability all stars bluer than $B-V \sim$0.40 mag in the HB region of the
CMD and did not find additional variables. Furthermore, in And~XIX as well as in other M31 dSphs 
we are  studying, we do find RRc stars, in spite of the contamination by background galaxies in that part of the CMD. 
Hence, we are inclined to think that the paucity of first overtone pulsators in And~XXI is a real feature.

\begin{figure}[!]
\includegraphics[width=7.7cm,clip]{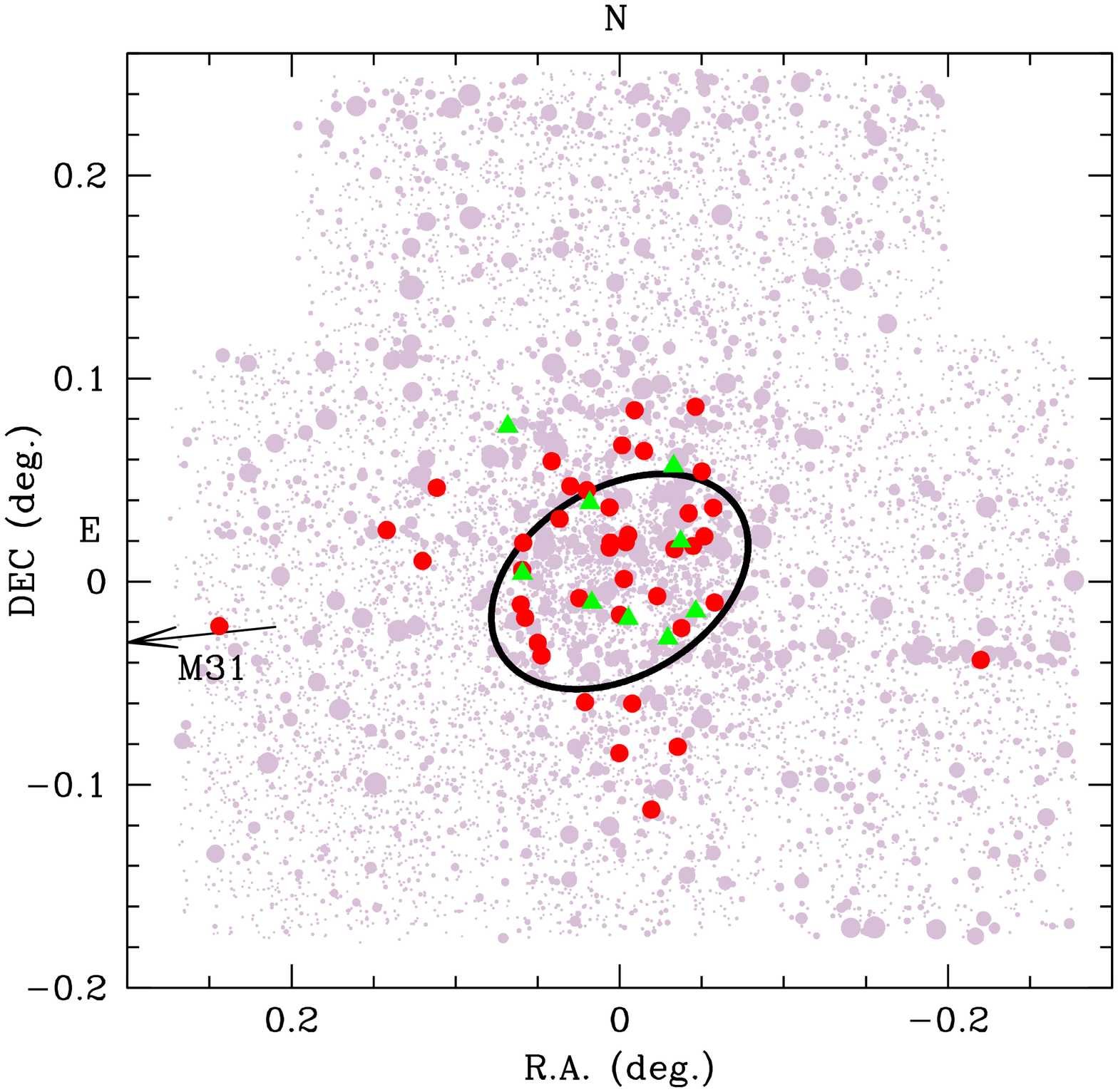}
\caption[]{Spatial distribution of the variable stars detected in And~XXI.  
Red filled circles mark the RR Lyrae stars 
and green filled triangles  are the 
ACs. Symbol size for all other stars is proportional to the magnitude. An ellipse marks the area enclosed in the half-light radius defined by Martin et al.(2009).
The arrow marks the direction towards the M31 center.}
\label{fig:spa}
\end{figure}

The spatial distribution of the RR Lyrae stars is shown in Figure \ref{fig:spa} (filled red circles), and indicates a rather complex structure for And~XXI.
The RR Lyrae stars seem to be asymmetrically distributed.  
In
particular,  outside the galaxy half-light radius we note two elongated features 
traced by the variable stars, one  $\sim$ 12 arcmin long going from North-West to South-East 
and the other pointing 
 towards South. They may trace past interactions with a dwarf satellite galaxy of M31 and/or with M31 itself. We discuss this more in detail in Section~\ref{sec:cmd}.

\subsection{Anomalous Cepheids}

We found in And~XXI nine variable stars which are $\sim$ 1 mag brighter than the average magnitude of the RR Lyrae stars.
We classified these stars as ACs based on the comparison with stellar isochrones 
  and the period-Wesenheit ($PW$) relation of  ACs.
The Wesenheit function \citep{van75,mad82}  is  defined as: W($B,V$)= M$_V -$ 3.1$\times$($B-V$), where  M$_V$ is the $V$ magnitude 
corrected for the distance. The $\langle V \rangle $ magnitudes of the ACs were corrected adopting the distance
modulus (m-M)$_0$=$24.40\pm0.17$ mag  derived from the RR Lyrae stars (see Section~\ref{sec:dist}) and used to derive the corresponding Wesenheit indices.
The $PW$ relations of  And~XXI ACs are shown in the left panel of Figure~\ref{fig:plac}  along  
with  the $PW$ relation for Fundamental (F) mode and First Overtone (FO) mode 
ACs recently derived for the    
Large Magellanic Cloud (LMC, solid lines) by  \citet{ripe14}\footnote{\citet{ripe14}'s relations were derived for the $V$ and $I$ bands, 
and we have  converted them to $B$ and $V$ 
 using equation 12 of \citet{mar04}.}. 
 The nine  ACs in And~XXI well follow, within the errors, the \citet{ripe14} relations 
 for F and FO pulsators. 
 To confirm that these bright variables are mostly ACs, in the right panel of Figure~\ref{fig:plac} we compare them with 
 the $PW$ relation for Classical Cepheids (CCs) in the LMC 
derived by \citet{sos08}. The $PW$ relations for 
CCs indeed do not fit very well the bright variables in 
And~XXI. However, we cannot totally reject the possibility that at least one, perhaps two of the bright variables (namely, V4 and V26) could be CCs.
As in Paper~I we derived the specific frequency of ACs in the galaxy, on the assumption that all the 9 bright variables in And~XXI are bona fide ACs. This is 
plotted in Figure~\ref{fig:spec} along with the AC specific frequency in  a number of MW and M31 dwarf satellites.
 And~XXI  (starred symbol in  Figure~\ref{fig:spec}) well follows the relation traced by the other dwarf galaxies. 
 The error is computed considering a Poisson statistics and include the case in which 2 of the ACs are in fact CCs.

\subsection{Comments on individual peculiar stars}\label{sec:comments}
V2 - The star is 0.3 mag fainter than the average $V$ magnitude of RR Lyrae stars.
The pulsation amplitude in the $V$ band  is  larger than  in the $B$ band 
 and the V light curve although well sampled is very noisy. The color of the star
is blue ($(B-V)=0.39$ mag) for an RRab. Only V28 that is an RRc star is bluer. V2 is likely  a 
RR Lyrae star of the M31 field.

V4 - The  $V$ light curve is sparsely sampled between phase 0.40 and 0.70.
However, maximum and minimum light are well covered, hence the  intensity-averaged mean $V$ magnitude 
should be correct.  The star is about 1.5 magnitude brighter than the horizontal branch and in the $PW$ plane V4 falls within  1 $\sigma$ 
from the relation for first overtone ACs, but also  fits  well the relation
for fundamental-mode CCs.

V6 -  Due to the star  pulsation period (P=1.067 d) and our data 
sampling, we do not have data points in the rising 
part of the light curve both in $B$ and $V$. This  affects 
 the star mean magnitudes. In fact 
in the $PW$ plane V6 is slightly  off  the relation for FO ACs.

\begin{figure*}[]
\begin{center}
\includegraphics[width=16.3cm,clip]{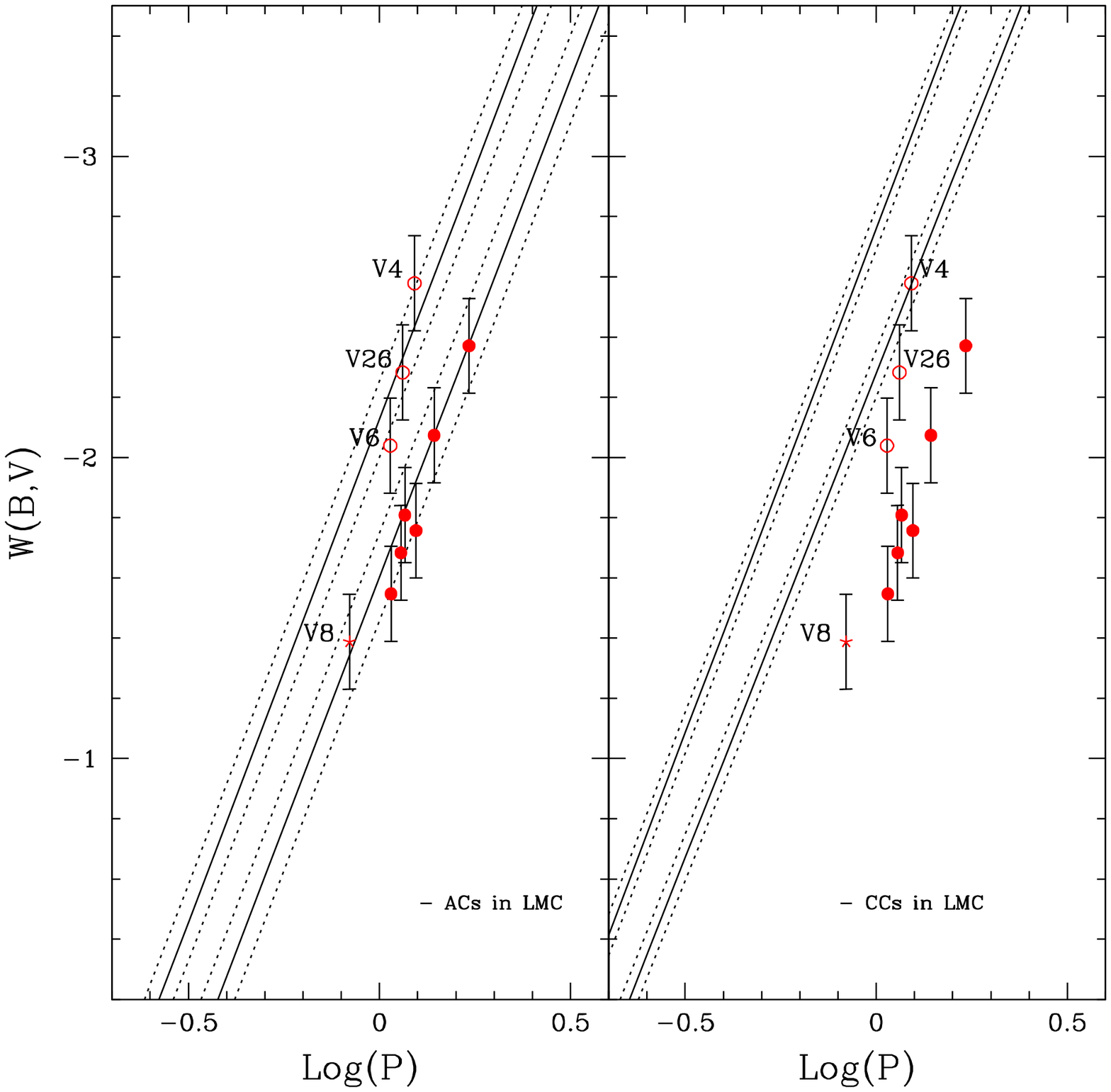}
\end{center}
\caption[]{$PW$ relations for the 9 variable stars in And~XXI which are brighter than the horizontal branch. Red filled circles are fundamental mode pulsators, 
open red circles are first-overtone variables. 
Also plotted as a star is the bright RR Lyrae variable V8  (see details in the text). 
Solid lines in the left panel are the $PW$ relations for ACs in the LMC (Ripepi et al. 2014), while on the right panel are shown the $PW$ relations for CCs in the LMC  from Soszynski et al. (2008). Dashed lines show 
the 1 $\sigma$ contours.}
\label{fig:plac}
\end{figure*}

\begin{figure*}[]
\includegraphics[width=16.3cm,clip]{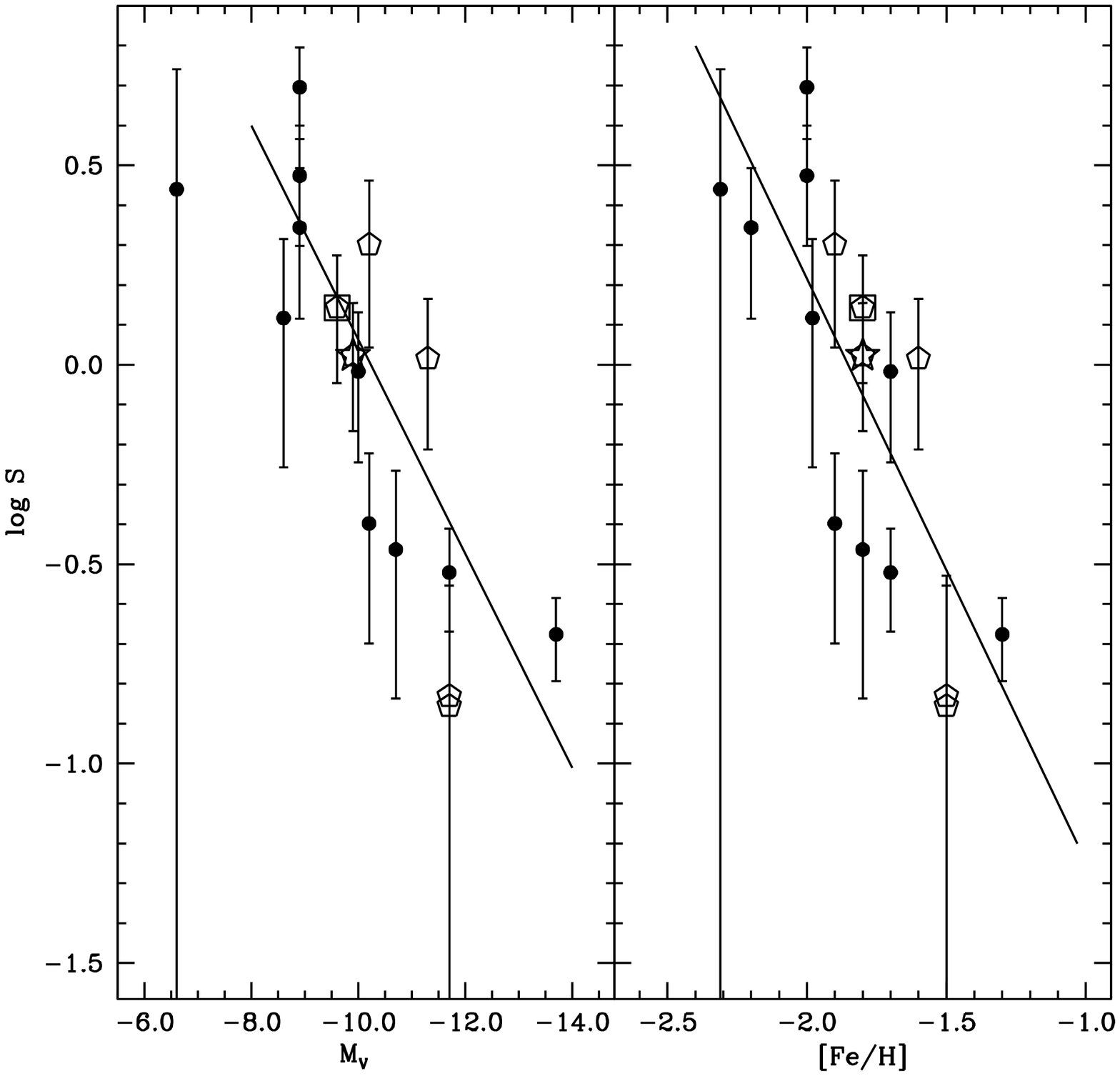}
\caption[]{$Left$: Specific frequency of ACs in dwarf satellites of the MW (filled circles)
and M31 (open pentagons) versus absolute visual magnitude. And~XXI is represented by a starred symbol, And~XIX by an open square.
 The black lines in both panels are the best fits to the data.
$Right$: Same as in the left panel, but versus metallicity of the parent galaxy.}
\label{fig:spec}
\end{figure*}

V8 - This is the longest period (P=0.8342 d) and the brightest ($V$=24.93 mag) RR Lyrae in our sample.  It has very similar amplitudes
 in $B$ and $V$. V8 is also the reddest variable in And~XXI.
  In Figure~\ref{fig:plac} V8 is plotted with a star sign, and shows to follow well the $PW$ relation for F-mode ACs, 
 suggesting that this star could be an AC.

V26 - As for  V6 this AC lacks part of the $B$ and $V$ light curves due to the combination
of pulsation period (P=1.1500 d)  and sparse data sampling.

V27 - This is one of the two brightest RR Lyrae stars in our sample. The $B$ and $V$ amplitudes  are small
for a RRab of that period (P=0.7052 days), nevertheless the amplitude ratio A$_{B}$/A$_{V}$=1.28 is typical of a RR Lyrae star. 

V46 - The star period is uncertain and the light curves are noisy. However amplitudes and average magnitudes are typical of a normal 
RR Lyrae star.

V49/V50- These two variables  are about  $12'$ away from the galaxy center. However, the pulsation period and the mean magnitudes 
are consistent with the average values of the RR Lyrae stars in And~XXI.  We will show
is Section~\ref{spatial} that portions  of And~XXI seem to extend that far from the galaxy center.


\section{DISTANCE}\label{sec:dist}
\begin{figure}[b!]
\includegraphics[width=8.0cm,clip]{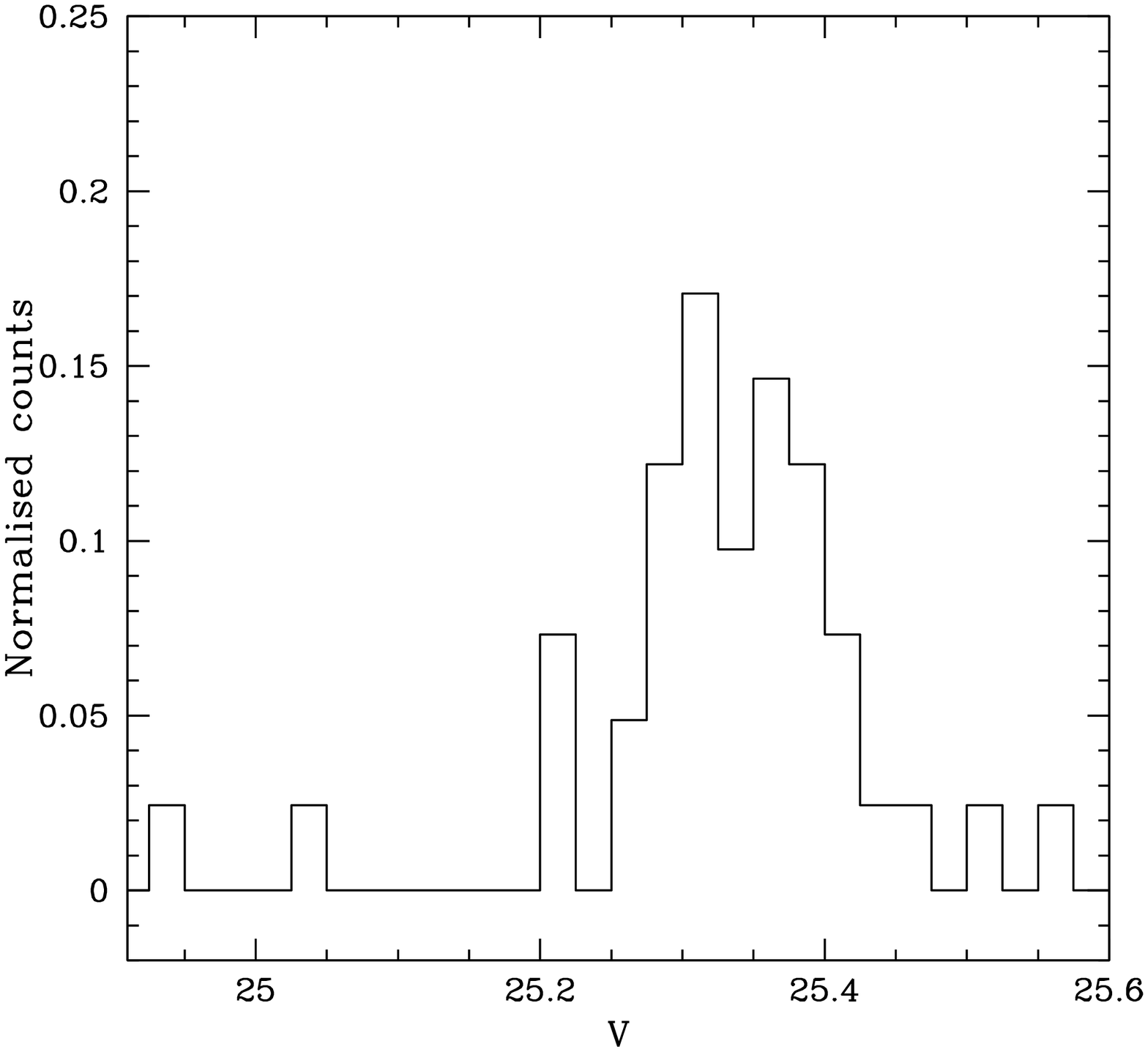}
\caption[]{Histogram of the intensity-averaged mean $V$ magnitudes  of  the RR Lyrae stars identified in And~XXI. 
The bin size is 0.025 mag. The two brightest stars are V8 and V27, the two faintest ones are V2 and V9.}
\label{fig:vmag}
\end{figure}

The distribution of the  intensity-averaged mean $V$ magnitudes
of the RR Lyrae stars in And~XXI is shown in the histogram of Figure~\ref{fig:vmag}.
The average magnitude of the 41 RR Lyrae stars in And~XXI is  $\langle V(RR) \rangle=25.33\pm0.11$ mag (average on 41 stars).
Considering only  RR Lyrae stars inside the area defined by the galaxy  r$_h$,  the average becomes 
$\langle V(RR) \rangle=25.32\pm0.13$ mag (average on 22 stars). 
 Among the RR Lyrae stars V8 and V27  are 0.3 mag brighter than the RR Lyrae average  
$V$ magnitude and are most probably blended objects or foreground RR Lyrae stars of the M31 halo. 
On the faint side variables V2 and V9 are $\sim$ 0.3 mag fainter than the RR Lyrae average  
$V$ magnitude and  possibly are RR Lyrae belonging either to the M31 halo or to a structure around M31.
Excluding these four stars (V2, V8, V9 and V27) 
the averages we obtain are $\langle V(RR) \rangle=25.33\pm0.06$ mag (average on 37 stars)  for all and 
$\langle V(RR) \rangle=25.32\pm0.12$ mag (average on 19 stars) for the RR Lyrae stars inside the galaxy r$_h$. 
We adopt the latter value for the average $V$ magnitude  of And~XXI RR Lyrae stars,  M$_{\rm V}=0.54\pm0.09$ mag  
for the absolute visual magnitude of RR Lyrae stars with metallicity of [Fe/H] = $-$1.5 dex \citep{cle03} and
$\frac{\Delta {\rm M_V}}{\Delta {\rm[Fe/H]}}=0.214\pm0.047$ mag/dex \citep{cle03,gra04} for the 
slope of the RR Lyrae luminosity-metallicity relation. 
For  And~XXI metallicity we adopt the value  [Fe/H]=$-1.8\pm0.4$ dex as derived spectroscopically by \citet{col13}.
To correct  for interstellar extinction we derived the reddening from the galaxy RR Lyrae stars  
using the method by \citet{pier02}.
This is based on the relation  between intrinsic $B-V$ color,  period,  amplitude
of the luminosity variation in the $B$ band and metallicity  of the RR Lyrae stars.
The reddening value we derive is $E(B-V)=0.15\pm0.04$ mag, assuming [Fe/H]=-1.8 dex for the metallicity.  
The $E(B-V)$ value and its r.m.s do not change if  stars V2, V8, V9 and V27 are excluded. The $E(B-V)$ derived from the RR Lyrae stars  
is larger than the value of $E(B-V)=0.094\pm0.026$ mag  derived by \citet{sch98}, but still consistent in 1 $\sigma$.
The distance modulus of And~XXI  derived from the RR Lyrae stars  using our reddening estimate is (m-M)$_0$=$24.40\pm0.17$  mag.
Our distance modulus is smaller than the one found by  Conn et al. [2012, (m-M)$_0$= $24.59^{+0.06}_{-0.07}$ mag] using the 
luminosity of the RGB tip,
although compatible within 1 $\sigma$. 
The difference  between the two distance estimates is mostly due to differences in the adopted reddening values.  \citet{con12}
use the reddening by \citet{sch98}   which is 0.057 mag smaller than our estimate from the RR Lyrae stars.
If one adopts the same reddening value the two distance moduli become identical.

\section{CMD AND STELLAR POPULATIONS}\label{sec:cmd}
\begin{figure*}[]
\centering
\includegraphics[scale=0.7]{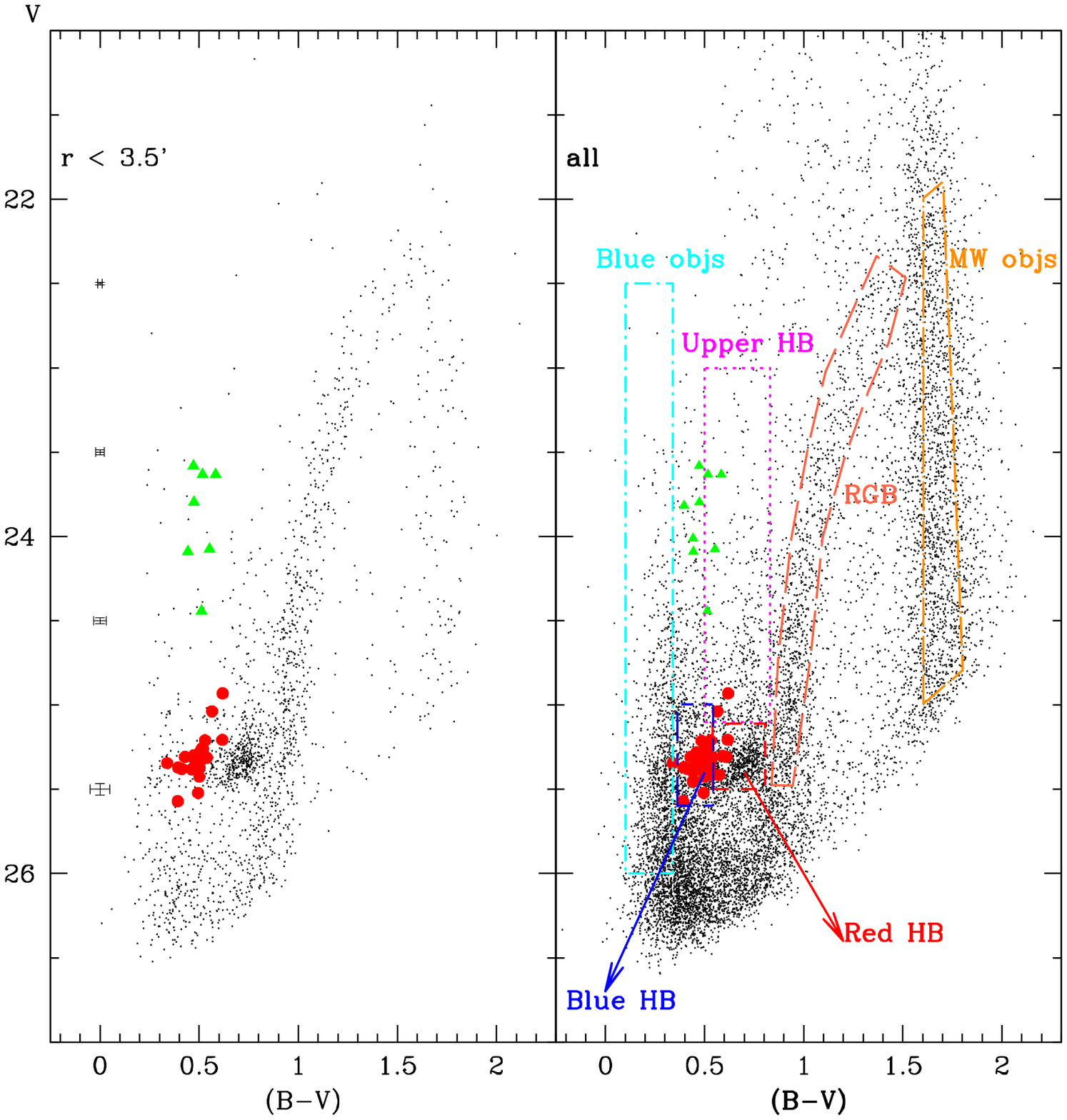}
\caption[]{$Left$: CMD of the sources in our photometric catalog with  $-0.3 \le$ Sharpness $\le 0.3$, $\chi < 1.1 $ and 
 located within the area delimited by the galaxy half-light radius and the ellipticity  by Martin et al. (2009). Red  
circles mark the RR Lyrae stars while the green triangles are ACs. Typical error bars are shown on the left. $Right$:
As in the left panel, but considering  sources in the whole LBC FOV. The regions in the CMD mark different
selections performed to study the projected distributions (see Section~\ref{spatial}). }
\label{fig:cmd}
\end{figure*}

\begin{figure*}[]
\centering
\includegraphics[width=13cm,height=16.5cm]{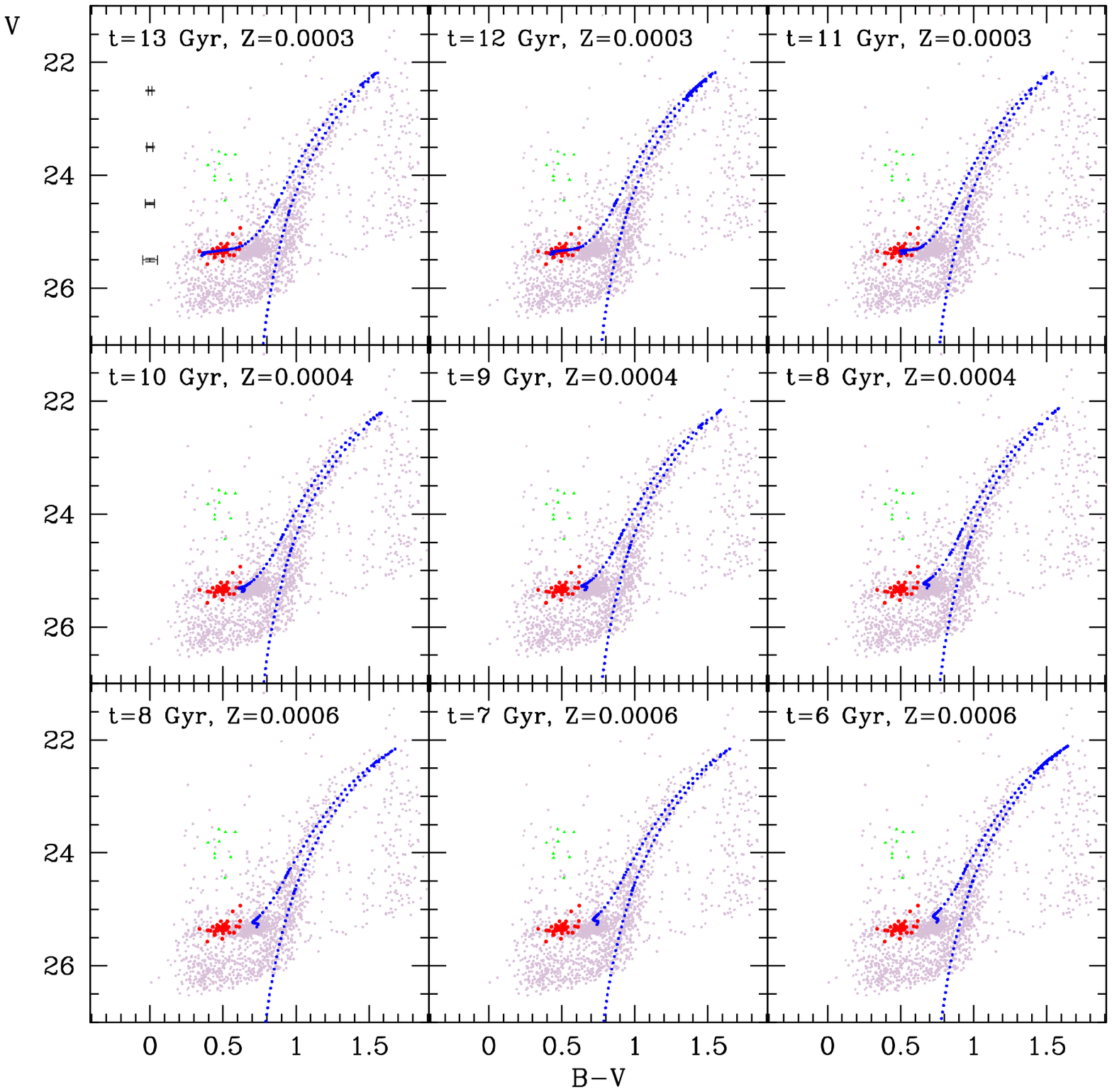}
\caption[]{CMD of And~XXI overlaid by Padova stellar isochrones (Bressan et al. 2012) with different age (13-6 Gyr) and 
metallicity  (Z=0.0003, Z=0.0004 and Z=0.0006; from top to bottom).  Red filled circles are RR Lyrae stars,  the green filled triangles are ACs.}
\label{fig:pana}
\end{figure*}

The CMD of And~XXI obtained in the present study  is shown in Figure~\ref{fig:cmd}, where  
in the left panel 
 we plotted only sources located 
within the area delimited by the galaxy half-light radius and the ellipticity by Martin et al. (2009), whereas the right panel shows the CMD of 
all sources in the LBC field of view (FOV). To mitigate contamination from background  galaxies and peculiar objects
we selected our photometric catalog using the $\chi$ and Sharpness quality parameters provided by \texttt{ALLFRAME} and only retained 
sources with $-0.3 \le$ Sharpness $\le 0.3$ and $\chi < 1.1 $.
The variable stars 
are  plotted with red filled circles and green triangles for RR Lyrae
stars and ACs, respectively.  Prominent features of the CMD are the blue horizontal branch (HB) traced by the RR Lyrae stars, the red HB and the red giant branch (RGB). 
The sequence of stars at $B-V\sim 1.5-1.7$ mag and extending blueward
around $V=22$ mag is mostly composed by foreground field stars.
  The relative occurrence of both RR Lyrae stars and ACs in And~XXI   suggests  that the galaxy hosts different stellar generations.
 The CMD features and the position of the variable stars in the CMD were compared to the Padova isochrones
obtained using 
the CMD 2.5 web interface ($http://stev.oapd.inaf.it/cgi-bin/cmd$) which is based on models from \citet{bres12}.  This is shown in 
Figure~\ref{fig:pana}. 
Isochrones from 11 to 13 Gyr and Z=0.0003 fit well the blue part of the RGB and the position of the RR Lyrae stars, (see top panels
of Figure~\ref{fig:pana}). \citet{bres12} use for the Sun a value of Z=0.0152 hence the tracks at Z=0.0003 correspond 
to [Fe/H]=-1.7 dex. The central and red part of the RGB together with red HB are well fitted by isochrones from 6 to 10 Gyr and 
enhanced metallicity ranging from Z=0.0004 to Z=0.0006 ([Fe/H]$\sim$-1.5 dex), as shown in central and bottom panels of Figure~\ref{fig:pana}.
The intermediate age population seems to be dominant in And~XXI given  
the best fit of the RGB with the 6-10 Gyr isochrones
and the high number of HB-red stars (692) compared to HB-blue stars (368), and  considering also that 
 the contamination by background unresolved galaxies
is much more prominent in the HB-blue. 
Another possible argument in favor of this hypothesis is 
the paucity of RRc stars. In the literature there are  
other systems with similar properties  of metallicity and average period of the RRab stars 
which totally  lack RRc stars, the Galactic globular clusters Ruprecht 106 \citep[Rup 106, ][]{kaluzny1995} and NGC 5824 \citep{meis2006}
 that both are younger than other Galactic globular clusters with same metallicity.

\begin{figure*}[]
\centering
\includegraphics[width=13cm,height=16.5cm]{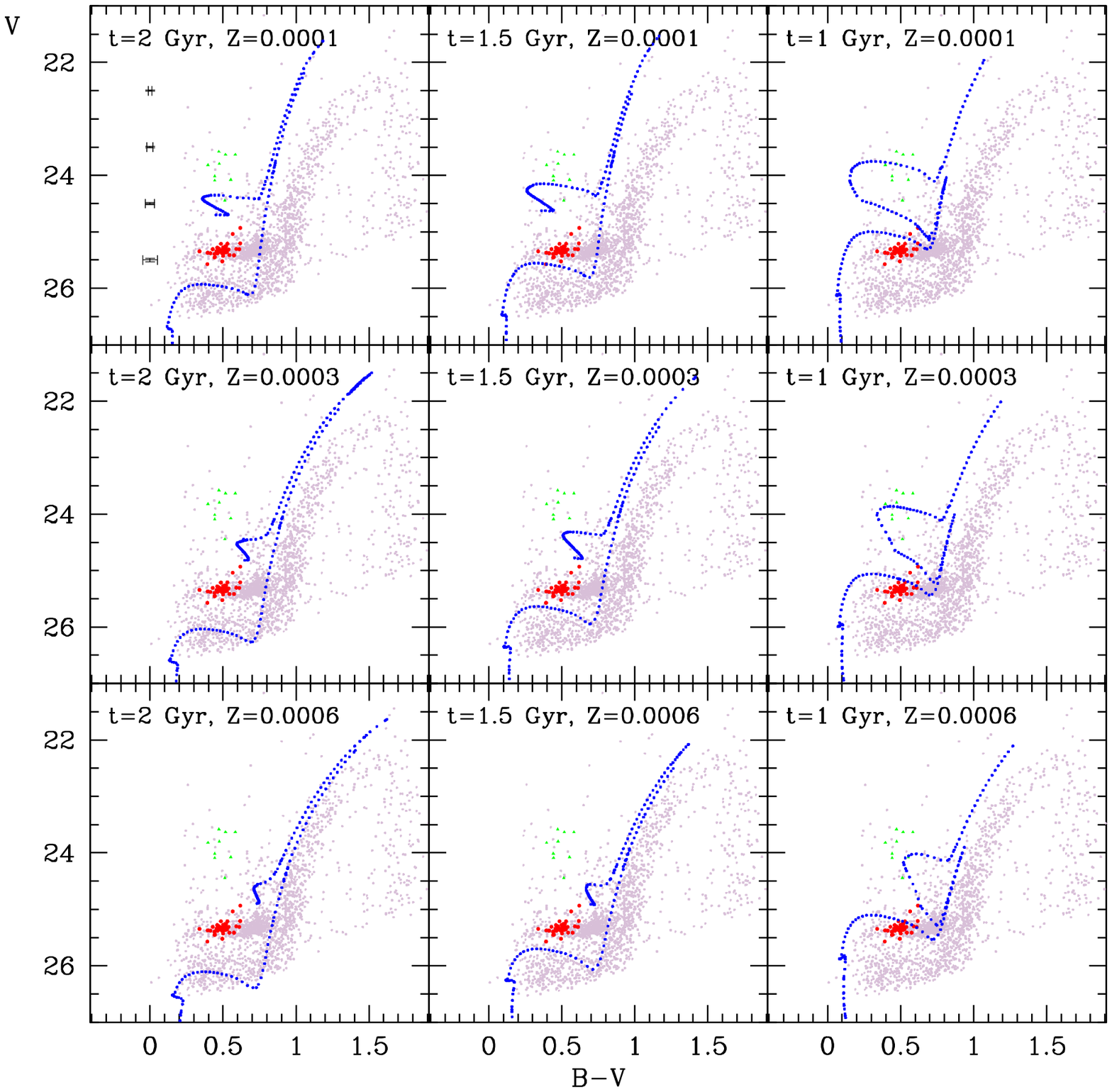}
\caption[]{Same as in Figure~\ref{fig:pana}, but for isochrones of 1, 1.5 and 2 Gyr and metallicity Z=0.0001, Z=0.0003 and Z=0.0006 (from top to bottom).}
\label{fig:pany}
\end{figure*}

 The presence in And~XXI of 9 ACs gives hints of a possible stellar population
with age in the range of 1-2 Gyr.  Indeed,  Figure~\ref{fig:pany} shows that the 
position of the ACs in the CMD is well fitted by isochrones from 1 to 2 Gyr with metallicity in the range Z=0.0001-0.0006.
Furthermore, above the  HB we observe a sequence of objects that likely belong to this 
young stellar component.
We should remind  that the origin of ACs is still a matter of debate in the literature, 
with the two most accepted scenarios being that they either represent
young single stars produced by a relatively recent episode of star formation,
or that they formed from mass transfer in binary systems as old as the other stars in the galaxy. 
Both channels may produce ACs, however, perhaps a distinction can be made between the few ACs
found in old stellar systems like globular clusters and ultra faint dwarf galaxies (UFDs) like Hercules (\citealt{mus12}), that likely originate 
from binaries, and the larger numbers of ACs found in galaxies with gas and 
recent star formation like Leo~T \citep{cle12} among the UFDs and And~XIX (Paper I) and And ~XXI among the M31 satellites, that likely 
are bona fide young stars.

\section{PROJECTED DISTRIBUTIONS}\label{spatial}
\begin{figure*}[]
\centering
\includegraphics[width=14cm,height=16.5cm]{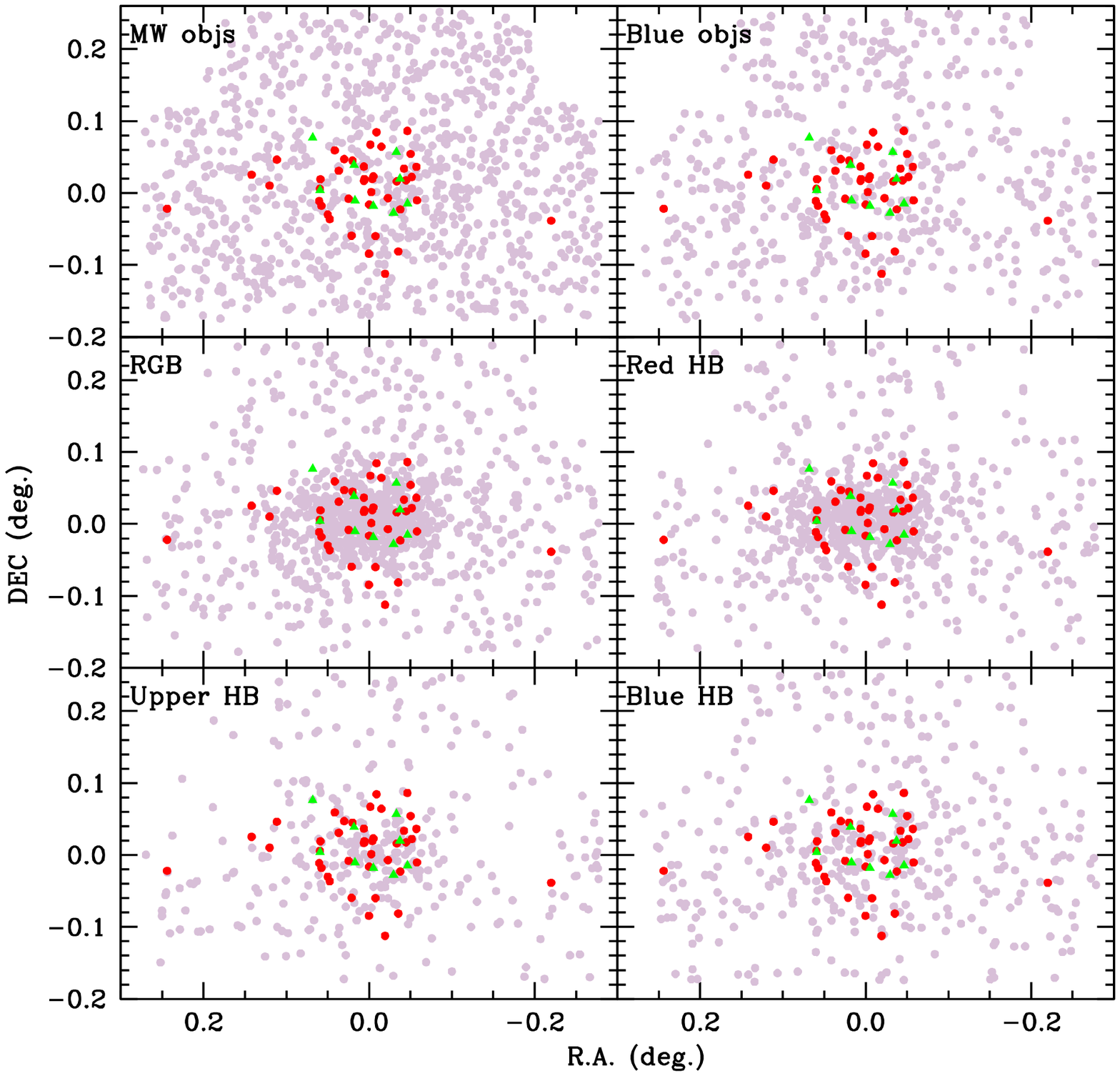}
\caption[]{ Projected distribution of selected samples of objects in the CMD of the total LBC FOV. 
Superimposed are the RR Lyrae stars (red, filled circles)
and the ACs (green, filled  triangles). }
\label{fig:dist}
\end{figure*}

To check whether the objects above the HB are young stars belonging to And~XXI and not contaminant sources,  we  
selected stars from the CMD of all the sources in the  LBC FOV  which have color in the  range $0.50<(B-V)<0.83$ mag and magnitude $23.0<V<25.1$ mag
(see right panel of Figure~\ref{fig:cmd} for the different selections). 
We plotted the selected objects in a RA and DEC map (see bottom left panel of Figure~\ref{fig:dist})  
and found evidence of a clustering  about $\sim$ 1 arcmin (0.017 deg) south-west from And~XXI  center.  
Although slightly offset with respect to the center of And~XXI these young objects  correlate with the general 
 projected distribution  of the CMD-selected RGB and red-HB stars, 
 as shown in Figure~\ref{fig:dist}, and likely are galaxy members. 
The distribution of RGB stars (middle-left panel of Figure~\ref{fig:dist}) is consistent with And~XXI  position angle and ellipticity as given by \citet{mart09},
while the red-HB stars (middle-right panel of Figure~\ref{fig:dist}) appear to be more concentrated toward the center than 
the RGB stars. That older stars have a more extended distribution compared to  younger populations is  a well known, general feature seen in dwarf galaxies (see e.g.  \citealt{Held2001,Tolstoy2004, Clementini2005, bel2014}).  

The blue objects between $0.1<(B-V)<0.4$ mag and $22.5<V< 26$ mag are most probably unresolved galaxies, 
as discussed in Paper~I. Their  projected distribution (top right panel in Figure~\ref{fig:dist}) is homogeneous over the whole LBC FOV, 
hence confirming that  they are contaminant sources.
The red objects between $1.5<(B-V)<2.1$ mag and $26<V<21$ mag are MW foreground stars  and are equally distributed all over the  
FOV (top left panel in Figure~\ref{fig:dist}).
Since there is a strong contamination from unresolved galaxies in the blue part of the CMD, it is a hard task to separate out 
 the blue-HB members of And~XXI. We selected  as blue-HB members  the stars falling in the CMD region  $0.35<(B-V)<0.65$ mag and $25.2<V< 25.7$ mag, 
as  defined by the RR Lyrae variables identified in this work.
In the  projected distribution, the  blue HB stars (bottom right panel in Figure~\ref{fig:dist})
are spread almost homogeneously over the galaxy and although the contamination from background
sources may be significant, the HB-selected stars follow the galaxy position angle.

\begin{figure*}[t!]
\centering
\includegraphics*[width=6cm,height=6.5cm]{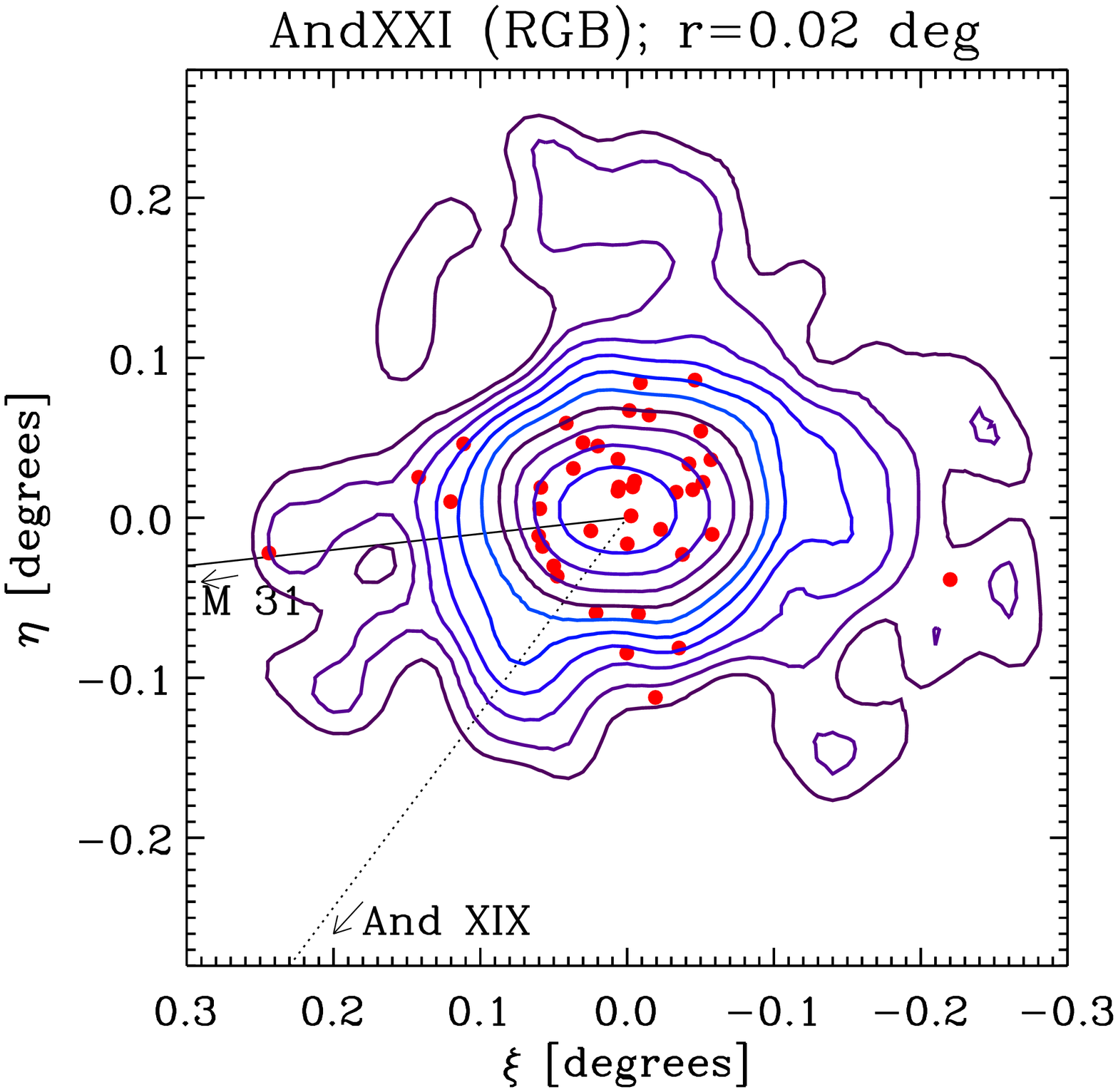}\includegraphics*[width=6cm,height=6.5cm]{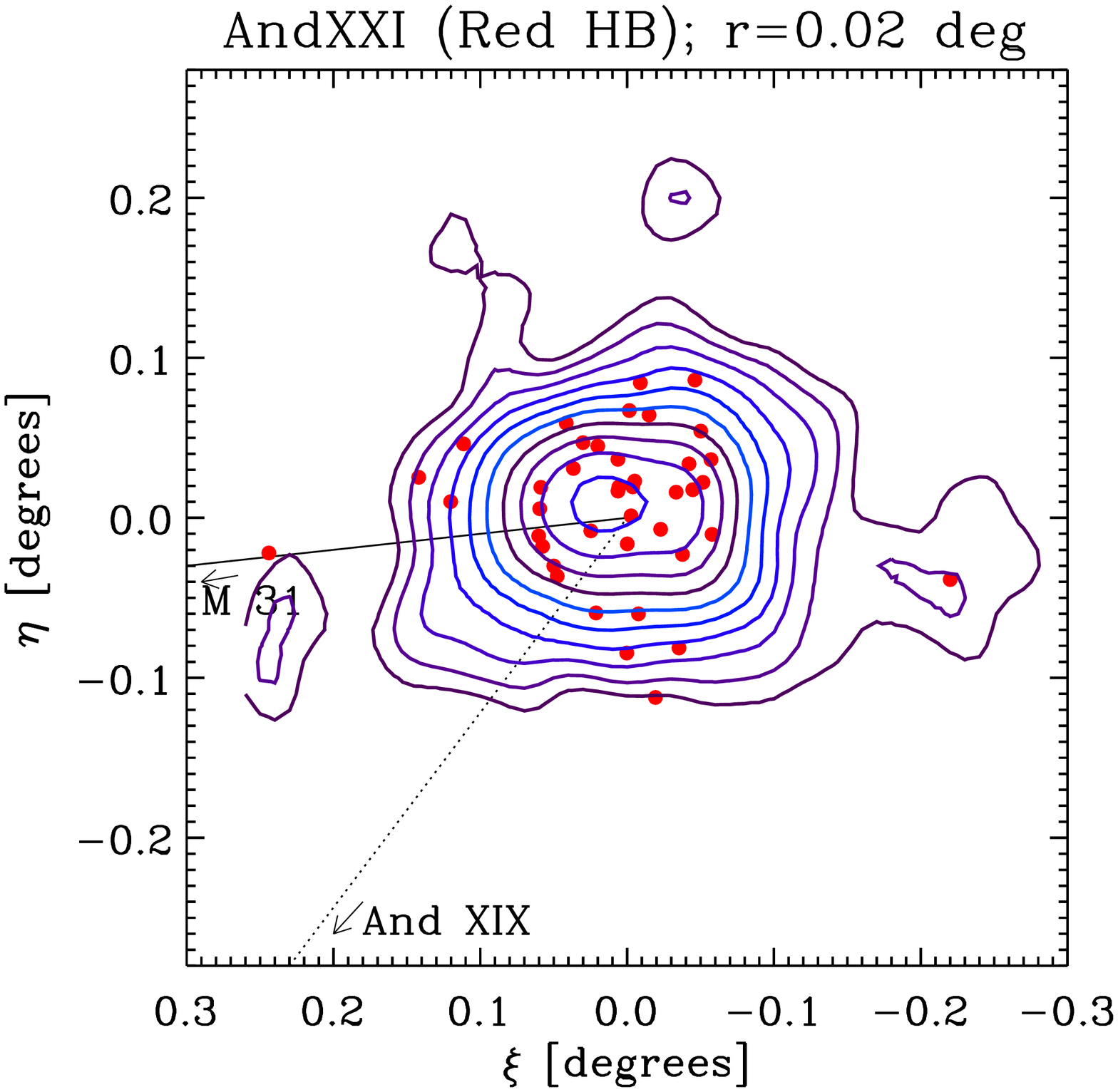}\includegraphics*[width=6cm,height=6.5cm]{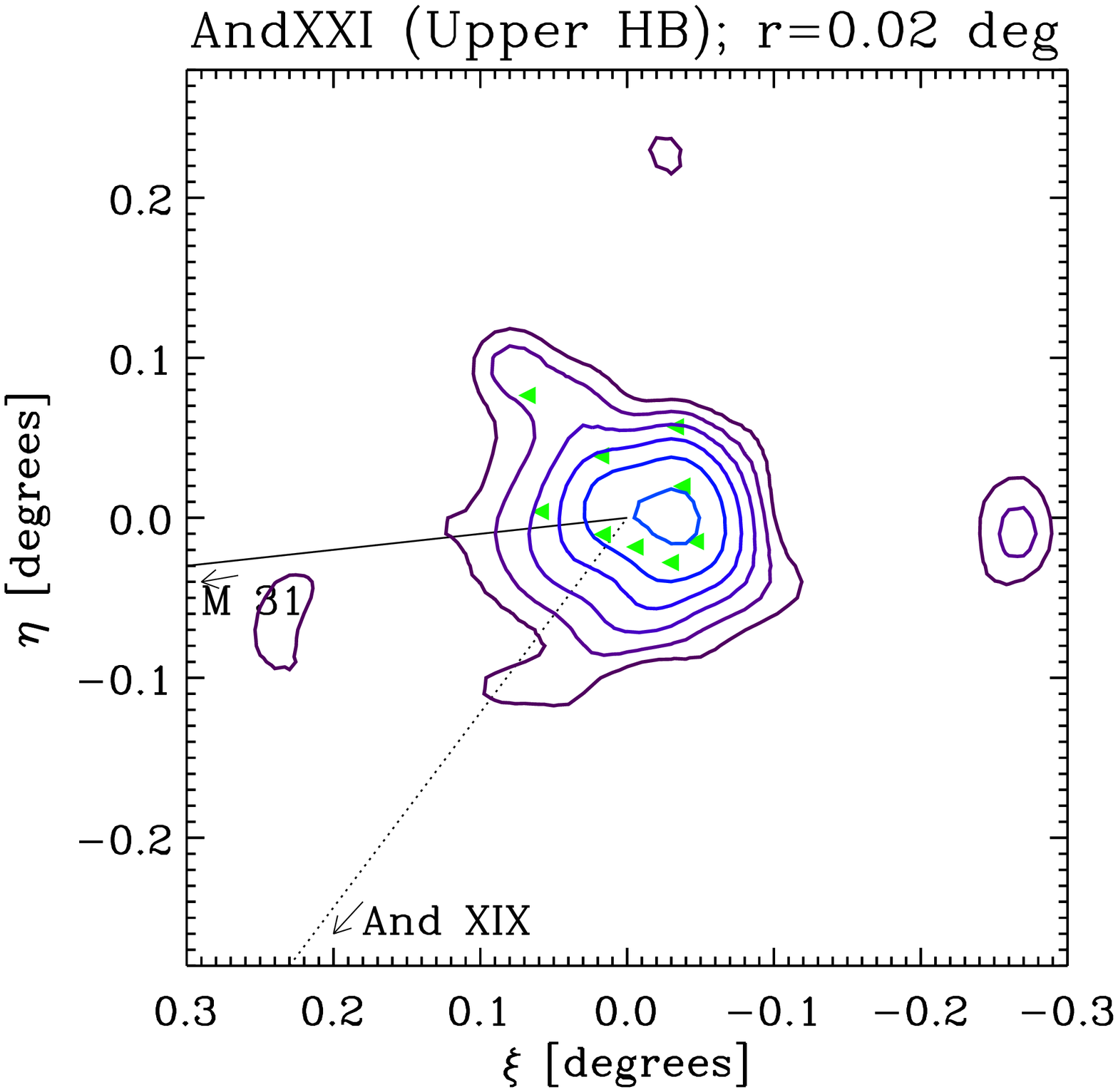}
\caption[]{Isodensity contours of three selected samples of stars in And~XXI. 
 $\xi$ and $\eta$ are the RA and DEC computed 
from the center of the galaxy given by Martin et al. (2009) and r is the FWHM of the Gaussian kernel. {\it Left panel:} RGB stars, {\it Center panel:} red-HB stars, 
{\it Right panel:}  upper-HB stars. RR Lyrae stars and ACs are also plotted as red filled circles and green filled triangles, respectively.}
\label{fig:isoden}
\end{figure*}

In order to identify particular structures in And~XXI, we built 
isodensity maps of the RGB, red-HB and  upper-HB stars. They are shown in Figure~\ref{fig:isoden}.
Stars were binned in $1.2'\times1.2'$ boxes and smoothed by a Gaussian
kernel of full width at half maximum (FWHM) of $1.2'$ (0.02 deg). The first contour levels are 3 $\sigma$ above
the sky background.
Directions to M31 and  And~XIX, which is the biggest satellite in the neighbourhood of And~XXI,
are also shown in  the figure. 
 The projected relative distance of And~XXI  to M31 and And~XIX is of $\sim 130$ kpc and $\sim $ 100 kpc, respectively.
The RGB stars are spread over the LBC FOV for more than twice the galaxy r$_h$. This 
was already clear for the presence of RR Lyrae stars far from the galaxy center.
Up to 6-7 arcmin ($\sim$ 0.11 deg) from the center the isodensity contours well follow the position angle given by  \citet{mart09}.
Right after 7 arcmin the isodensities appear to be twisted and in the east part of the LBC FOV are distorted 
in the direction of And~XIX. A sort of arm highlighted by the presence of one RR Lyrae star 
extends in the direction of M31.
The contours of red-HB and RGB stars have a similar distribution, while the  upper-HB stars
have different center and orientation when compared to the other two populations. The three populations
 show overdensities in the east, north and west directions. Two RR Lyrae stars are located  at the position
of the east and west overdensities, thus endorsing the presence of sub-structures
far from And~XXI center.

\begin{figure}[]
\centering
\includegraphics[width=8cm,height=8cm]{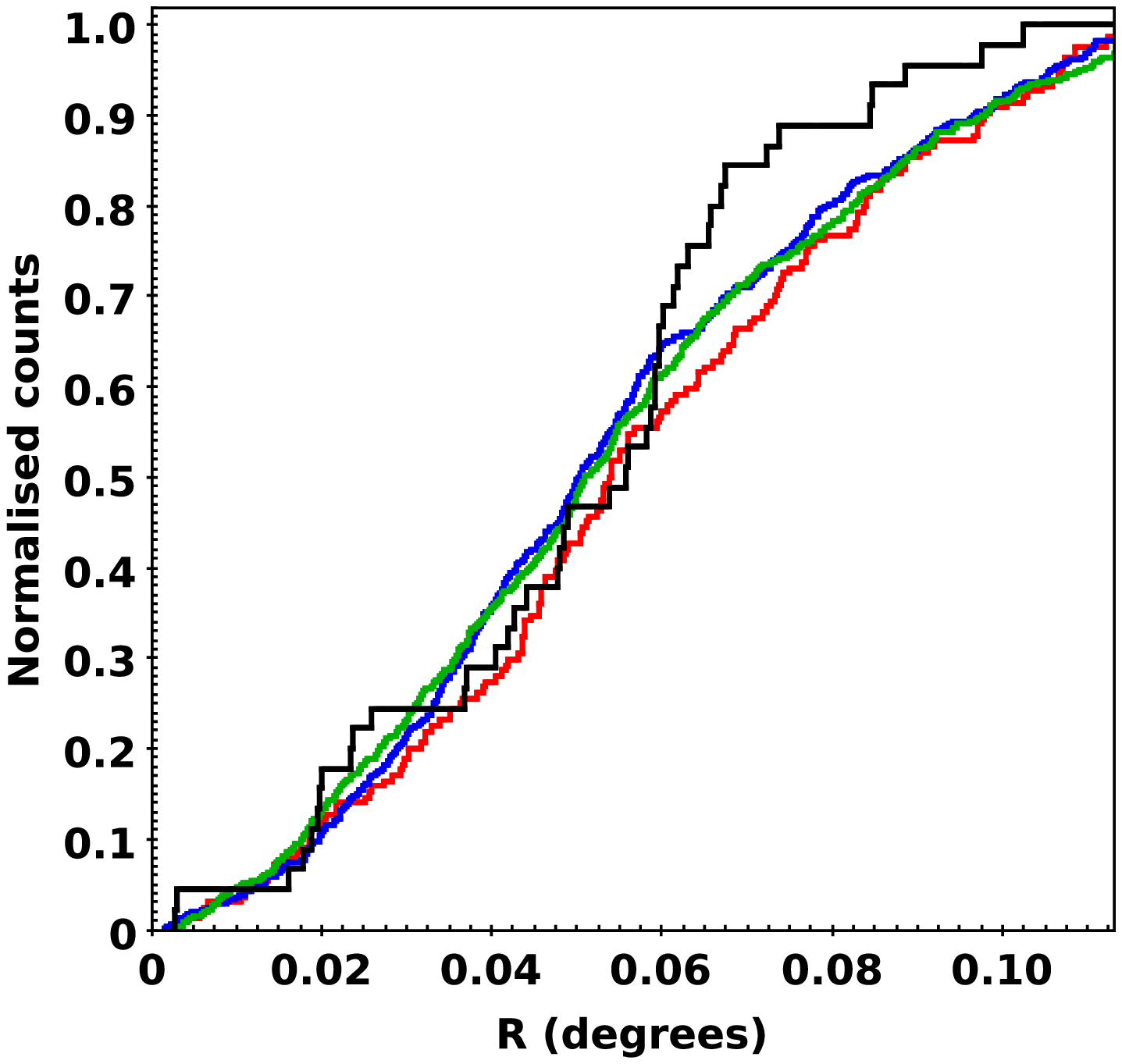}
\caption[]{Cumulative  radial distributions of selected samples of sources  in the CMD of And~XXI.
Green line: cumulative distribution for the RGB stars, blue line:  distribution of the red-HB stars,  red line:  distribution of the upper-HB sources,  
black line: distribution of the variable stars.}
\label{fig:rad}
\end{figure}

Figure~\ref{fig:rad} shows the cumulative radial distributions of 
selected samples of stars located inside twice the r$_h$ of  And~XXI.
We performed a two-sample Kolmogorov-Smirnov (K-S) test between these radial distributions.
The RGB (green line) and red-HB star (blue line) distributions shown in Figure~\ref{fig:rad} 
are indistinguishable  and likely arise 
from the same stellar generation. In fact,  the  K-S test provides a p-value as large as p=0.84, 
whereas the  K-S test  between  upper-HB population (red line) and  RGB stars  
gives a p-value of only p=0.24. 

Finally, though the number of variables  (RRLs+ ACs) is very small when compared 
to the number RGB stars the K-S test gives p=0.49 between the two samples.

\section{MERGING?}

In Section~\ref{sec:rrli} we showed the presence of two peaks in the period distribution
of And~XXI RRab stars. The double peak in the period distribution can be  a clue for the presence of two separate populations of RR Lyrae
stars, with slightly different age and/or
metallicity. Selecting stars with period within 1 $\sigma$ from the two peaks in period, 
we find an average magnitude of $V$=$25.35\pm0.08$ mag 
for stars with P$_1$=0.60 ($\sigma=0.02$) d and $V$=$25.25\pm0.10$ mag for  stars with P$_2$=0.68 ($\sigma=0.03$) d.
Although  the 0.1 mag difference between P$_1$ and P$_2$ RR Lyrae stars  is within the range of errors,
it goes in the direction that the longer period RR Lyrae stars (P$_2$) are brighter than the 
shorted period (P$_1$) ones, possibly due to a lower metal content.
 A difference in metallicity of ~0.4 dex can explain such  magnitude difference
according to 
\citet{cle03} and \citet{gra04} 
slope of the luminosity-metallicity relation of RR Lyrae stars. 
Since the RR Lyrae stars are produced by a population  older than 10 Gyr, the double-peaked period distribution 
might suggest that  either the 
primordial environment of And~XXI was enriched in only 2 Gyr or that the galaxy RR Lyrae stars arise from two different small 
galaxies that merged to form And~XXI.

\begin{figure*}[]
\centering
\includegraphics[scale=0.7]{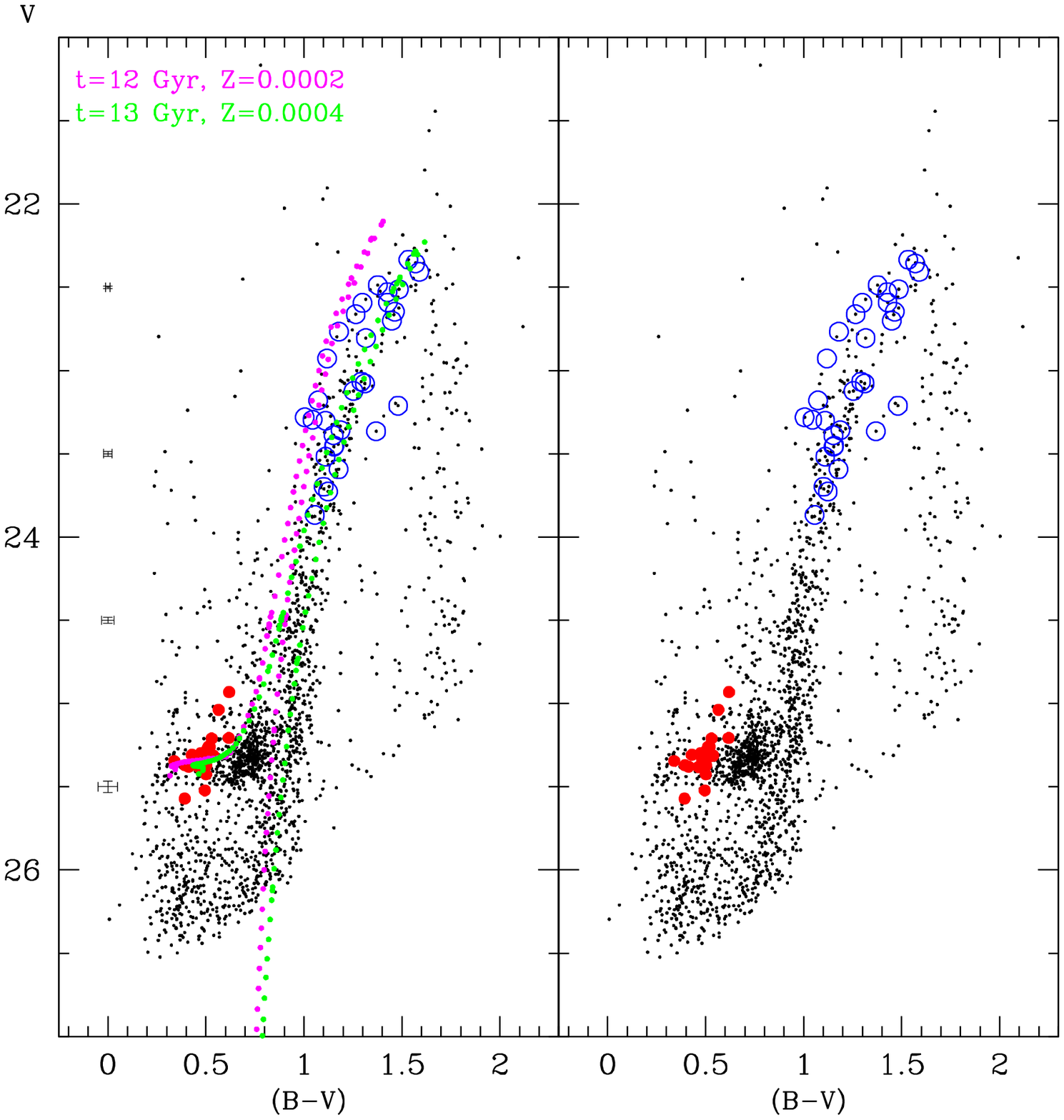}
\caption[]{CMD of And~XXI with marked as open blue circles  stars with membership spectroscopically confirmed by Collins et al.(2013).
Dashed lines are the Bressan et al.'s isochrones for 12 Gyr, Z=0.0002 (magenta line) and 13 Gyr, Z=0.0004 (green line), respectively.
Red filled circles are RR Lyrae stars. Typical photometric error bars are shown on the left.}
\label{fig:memb}
\end{figure*}

\begin{figure}[!]
\includegraphics[width=8cm,height=6cm]{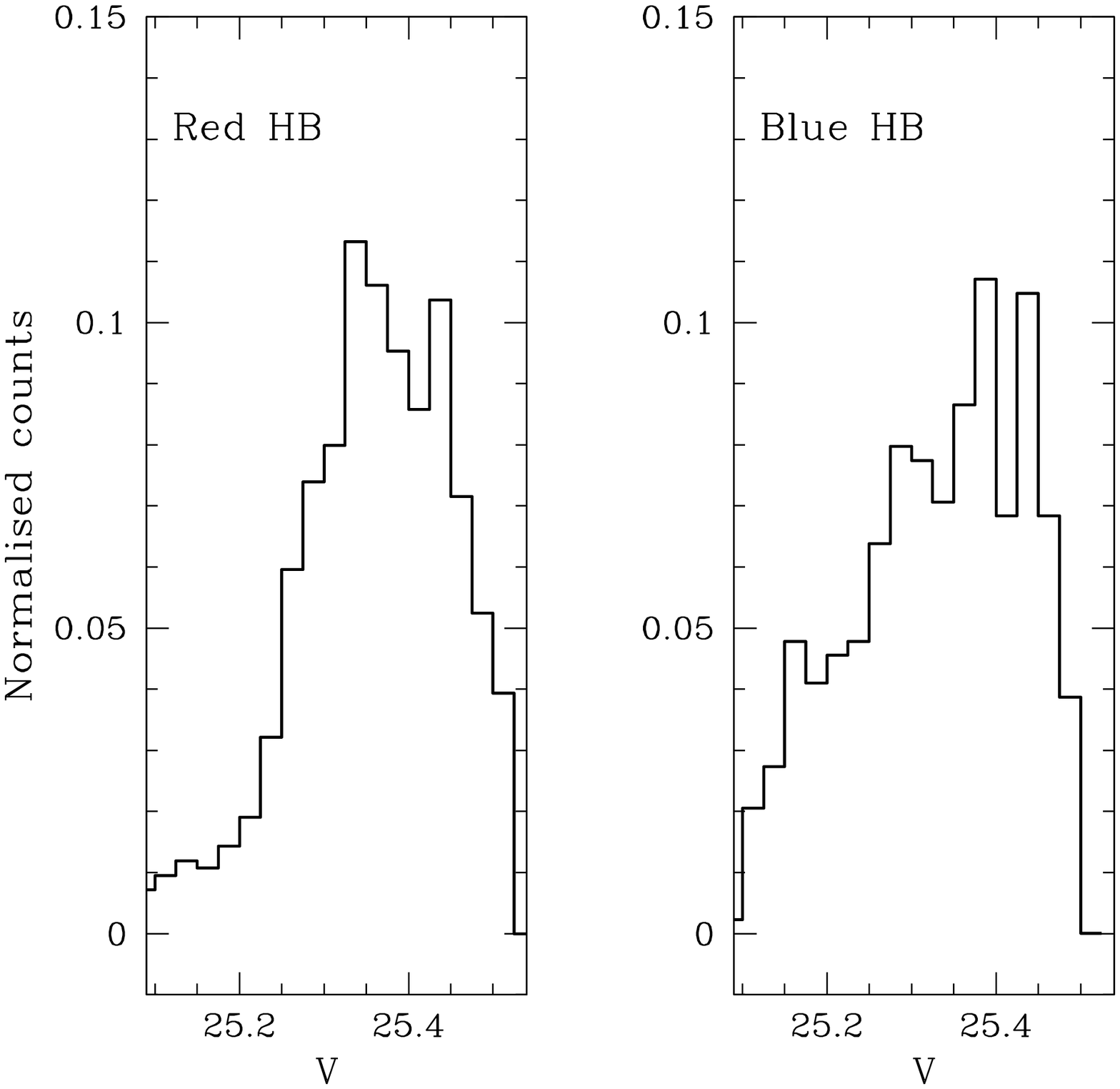}
\caption[]{Histogram of the $V$ magnitudes for the stars in the Red (left panel) and Blue (right panel) HB.}
\label{fig:vhis}
\end{figure}

Furthermore,  the RGB of And~XXI seems to be bifurcated, giving a further hint for the presence of two slightly different old populations.
To show that the spread of And~XXI  RGB is real and not due to contamination or photometric errors, we matched
our catalog with the  list of spectroscopic members of And~XXI identified by \citet{col13}. The latter 
are marked by blue open circles in the CMD of Figure~\ref{fig:memb}. They are mostly RGB stars 
that exhibit  a wide spread in color, at a given luminosity, due to age and/or metallicity differences.
As shown in the left panel of Figure~\ref{fig:memb} the member stars are enclosed within  the 12 Gyr, Z=0.0002 and 
the 13 Gyr, Z=0.0004 tracks. 
In addition the red-HB of And~XXI also separates  in two 
clumps with a difference of $\sim$ 0.15 in $V$ magnitude (see left panel of  Figure~\ref{fig:vhis}). 
A further argument in support of the merging scenario is the different  projected distributions of the selected
samples of stars in And~XXI. 

As early suggested by \citet{cole2004}  and later supported by \citet{yoz2012} with numerical simulations, 
the MW  dSph galaxy Fornax resulted from merging of two dwarf galaxies.
The visible remnant of this collision was detected by \citet{cole2004} as a small overdensity of young stars forming a shell structure 
displaced from Fornax center.
The young  stars in  And~XXI  (see Section~\ref{sec:cmd}) are indeed located in an aggregation
displaced from the center of the galaxy, which 
resembles the overdensity 
found in Fornax by \citet{cole2004}. Furthermore, And~XXI  variable stars seem to be placed
in separate shells around the galaxy center (see Figure~\ref{fig:spa}). These features suggest   also for And~XXI a merging scenario.

\begin{figure*}[]
\centering
\includegraphics[scale=0.7]{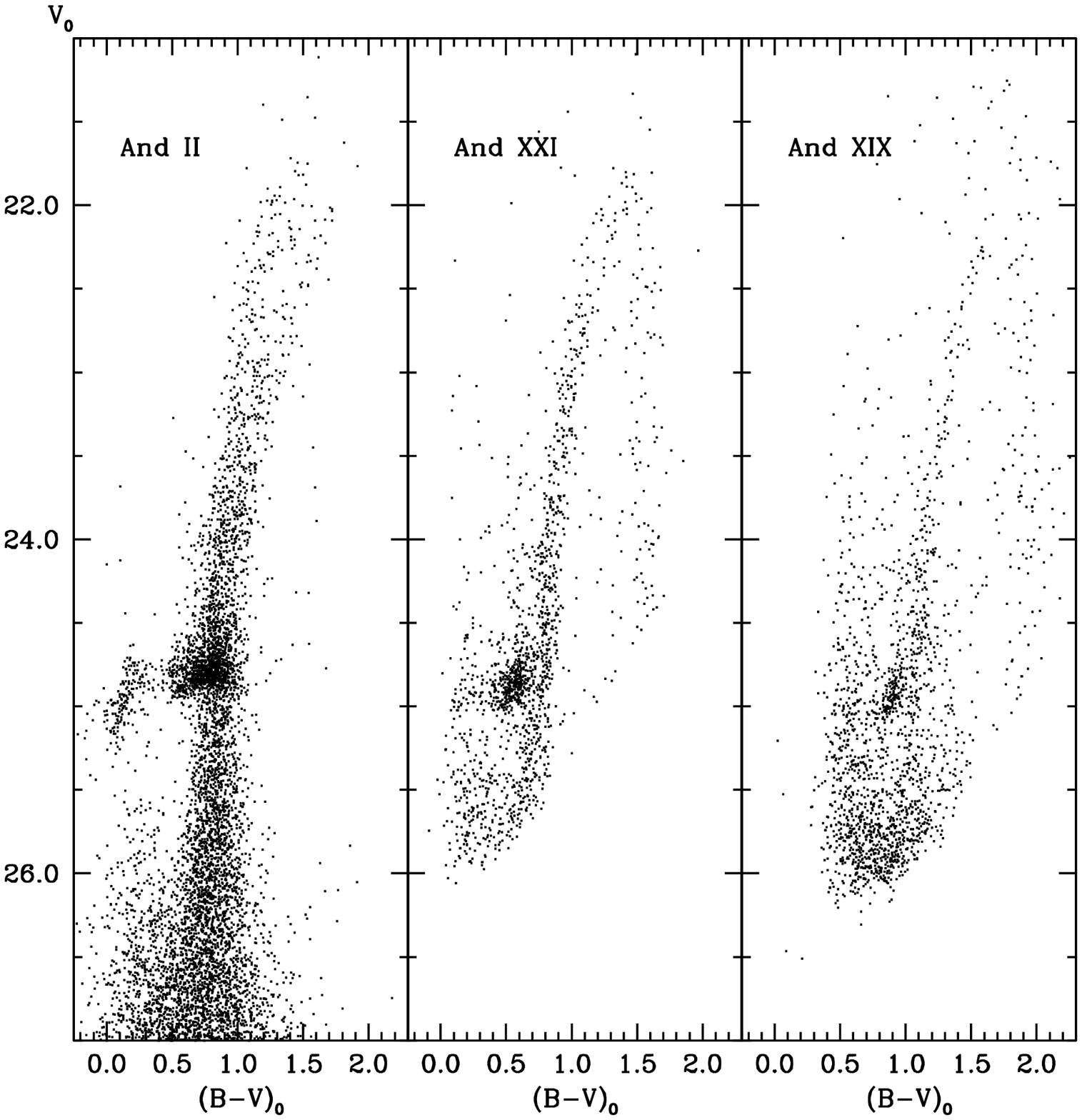}
\caption[]{Comparison of the de-reddened CMD of And~II (left panel) we obtained from HST archive data, with the CMDs of  And~XXI (central panel, this work)
and And~XIX (Paper~I). Only stars within the respective half-light radii are shown for And~XXI and And~XIX.}
\label{fig:comp}
\end{figure*}

According to \citet{amo2014}, among the M31 satellites,  And~II is  another example of 
dwarf  merging remnant. This would  explain the stellar stream they  detected kinematically  in  the galaxy.
%
The presence of two different old populations in  And~II is  supported also by 
the analysis of the CMD and by the properties of the variable stars.
The RR Lyrae stars of And~II were studied  by \citet{pri04}. Using  the periods in Table~2 of \citet{pri04},
we performed a  multi-gaussian fit 
and  found two separate peaks, 
 a first  
one at P=0.52 d with $\sigma$= 0.05 d and a second one at P=0.68 d with $\sigma$=0.01 d.
Hence, similarly to And~XXI, also And~II seems to host two distinct old stellar populations with different metallicity/age and corresponding different RR Lyrae populations, likely belonged to the two galaxies that merged in the past.
The left panel of Figure~\ref{fig:comp} shows  the CMD of And~II we obtained using  HST archive data (Prop ID 6514, PI: Gary Da Costa). 
The galaxy RGB  clearly bifurcates in two arms and  blue and red HBs also  seem to split.
CMD and variable stars  in And~II show properties  very similar  to And~XXI  (central panel of Fig.~\ref{fig:comp}).
For comparison the right panel of Fig.~\ref{fig:comp} shows the de-reddened CMD of And~XIX (Paper~I) inside the galaxy r$_h$ of 6.2 arcmin. 
The r$_h$ of And~XIX is approximately twice that of And~XXI and the  
contamination by foreground and background objects is much  higher. 
Still the RGB of And XIX is much narrower than 
the RGB of And XXI and And II.
In conclusion, we  speculatively attribute some of the  features of the CMD and RR Lyrae population of And~XXI
to the presence of two slightly different stellar populations belonging to two dwarf galaxies that merged in the past. 

Recently,  \citet{dea2014} investigated the frequency of merging between dwarf galaxies in the LG using the ELVIS simulations.
They found that the frequency  of satellite-satellite merging in the host virial radius of the MW and M31 is of the order of 10\%
and that this frequency doubles for satellites outside the virial radius. Considering the large number of satellites surrounding the MW and M31 
these predictions are consistent with both And~II and And~XXI being the result of satellite merging.

\section{CONCLUSIONS}
We have discovered a total of 50 variable stars in And~XXI, of which 41 are RR Lyrae stars and 9 are ACs. 
From the average period of the RRab stars $\langle$P$_{\rm ab}\rangle$=0.64 d
and the Bailey diagram we classify And~XXI as an Oo-II/Int object.
Using the variable stars and the CMD we found evidence  
for the presence in And~XXI of two major stellar populations,  a first one 12 Gyr old with [Fe/H]=-1.7 dex and 
a second, more conspicuous one with  an age of 6-10 Gyr  and  [Fe/H]=-1.5 dex. 
Furthermore, the discovery of 9 ACs traces also the presence in And~XXI  of a stellar population as young as 1-2 Gyr. 
This is similar to what we found in And~XIX (Paper~I), but  in And~XXI the 6-10 Gyr stellar component 
is much more dominant. Other dwarf  satellites of M31 contain RR Lyrae stars as well as prominent red clumps in the CMD 
\citep[e.g. see the CMD of And~I and And~III in Fig.1 by][]{pri05}.  
\citet{manc08} found that the population of And~V is mostly  composed of stars 8-10 Gyr old. 
Our results on the stellar population of And~XXI and And~XIX together with the above further literature evidences 
seem to suggest that a global event triggered star formation in the dwarf galaxy satellites of M31 about 6-10 Gyr ago.
An engine triggering this common star formation episode could possibly be the fly-by encounter
of the MW and M31 reported by \citet{zhao2013}

The  projected distribution of sources properly selected from the CMD of And~XXI  shows  the existence of an overdensity 
of young/intermediate age stars in a region
slightly off from the galaxy center. 
We discussed this evidence in light of  And~XXI possibly being the result of a merging
between two dwarf galaxies. This hypothesis is corroborated by the possible presence in the galaxy 
of two different RR Lyrae  populations as well as peculiar features in the CMD  (wide RGB, bifurcated  red HB) and from the  
presence of shell-like structures. If future spectroscopic observations will confirm And~XXI  to be a merged galaxy, 
 it would  become  the second such a satellite in M31,  
after And~II. 

\acknowledgments

We warmly thank P. Montegriffo for the development and maintenance of the GRATIS software,
G. Battaglia for providing the software 
to compute And~XXI's  density maps and A. Veropalumbo for software assistance.
Financial support for this research was provided by  PRIN INAF 2010 (PI: G. Clementini) and by Premiale LBT 2013. 
The LBT is an international collaboration among institutions in the United States, 
Italy, and Germany. LBT Corporation partners are The University of Arizona on behalf of the Arizona university system; 
Istituto Nazionale di Astrofisica, Italy; LBT Beteiligungsgesellschaft, Germany, representing 
the Max-Planck Society, the Astrophysical Institute Potsdam, and Heidelberg University; 
The Ohio State University; and The Research Corporation, on behalf of The University of Notre Dame, 
University of Minnesota, and University of Virginia. We acknowledge the support from the LBT-Italian 
Coordination Facility for the execution of observations, data distribution, and reduction.
Facility: LBT

\end{document}